\documentclass[aps,pre,reprint,amsmath,amssymb,showpacs,floatfix,longbibliography]{revtex4-1}
\usepackage{graphicx}
\usepackage{bm}
\usepackage{color}
\usepackage{algpseudocode}
\usepackage{algorithm}
\usepackage{multirow}
\usepackage{appendix}

\newcommand{\op}[1]{\hat{#1}}
\newcommand{\opsigma}{\hat{\sigma}}
\newcommand{\opH}{\hat{H}}

\begin{document}

\title{Guide to Exact Diagonalization Study of Quantum Thermalization}

\author{Jung-Hoon Jung}
\affiliation{Department of Physics, University of Seoul, Seoul 02504, Korea}
\author{Jae Dong Noh}
\email{jdnoh@uos.ac.kr}
\affiliation{Department of Physics, University of Seoul, Seoul 02504, Korea}
\date{\today}

\begin{abstract}
Exact diagonalization is a powerful numerical method to study isolated
quantum many-body systems. 
This paper provides a review of numerical algorithms to
diagonalize the Hamiltonian matrix.
Symmetry and the conservation law help us perform the numerical study 
efficiently. We explain the method to block-diagonalize the Hamiltonian matrix by
using particle number conservation, translational symmetry, 
particle-hole symmetry, and spatial reflection symmetry in the context of 
the spin-1/2 XXZ model or the hard-core boson model in a one-dimensional 
lattice. We also explain the method to study the unitary time evolution
governed by the Schr\"odinger equation and to calculate thermodynamic
quantities such as the entanglement entropy. As an application, we demonstrate
numerical results that support that the eigenstate thermalization hypothesis 
holds in the XXZ model.
\end{abstract}


\maketitle

\section{Introduction}\label{sec:intro}
Matrix diagonalization is one of the key numerical methods for solving physics problems.
From the eigenspectrum of relevant matrices, one can obtain the
normal modes of mechanical systems, the 
free energy of systems in thermal equilibrium, the probability
distribution of master equation systems, the energy levels of quantum
mechanical systems, and so on. 
Recently, quantum thermalization emerged as an interesting research topic
in the field of statistical physics~\cite{DAlessio:2016gr,Deutsch:2018fy}. 
It addresses the
question of how isolated quantum mechanical 
systems evolve into a state in thermal
equilibrium through a unitary time evolution. 
The eigenstate thermalization 
hypothesis~(ETH)~\cite{Srednicki:1996kn,Srednicki:1999bo} was 
suggested as an underlying mechanism for quantum thermalization and
has been attracting extensive theoretical~\cite{Rigol:2008bf} and 
experimental~\cite{Kaufman:2016jm,Abanin:2019dl} attention.
Theoretical works rely heavily 
on the computational method diagonalizing the Hamiltonian matrix.
Therefore, this presentation of a thorough 
review of computational techniques is timely. 
This article is a practical and pedagogical guide for the computational
method to obtain the whole eigenspectrum of the Hamiltonian and to
investigate thermodynamic properties of isolated quantum mechanical 
systems~(see also Ref.~\cite{Sandvik:2010kc}).

The Hilbert space dimension grows exponentially with the number of 
degrees of freedom. Thus, exploiting the symmetry property and the
conservation law through which the Hamiltonian matrix can be 
block-diagonalized is crucial. 
We will explain the block-diagonalization method and other useful numerical
algorithms by using the explicit example of the spin-1/2 XXZ Hamiltonian 
in a one-dimensional lattice. It is mapped to the spinless fermion system via
the Wigner-Jordan transformation~\cite{Izyumov:1988vi}.
It is also mapped to the hard-core boson model~\cite{Cazalilla:2011dm}. 
The model
is one of the best studied systems because of its simplicity and relevance to
experimental systems such as a system of the ultracold atoms~\cite{Kaufman:2016jm,
Rigol:2009ew, Santos:2010gk, Kim:2014kw,Garrison:2018hi, Steinigeweg:2013dc,
Noh:2019gx, Yoshizawa:2018js}.

This paper is organized as follows:
In Section~\ref{sec:model}, we introduce the XXZ Hamiltonian and the hard-core
boson Hamiltonian. We introduce the occupation number representation of
the basis states and explain a method to construct the Hamiltonian matrix.
In Section~\ref{sec:symmetry}, we explain the method to construct the
block-diagonal form of the Hamiltonian matrix by using 
the conservation law and the symmetry
property. In Section~\ref{sec:nd}, we present the performance of the numerical
algorithm. In Section~\ref{sec:thermal}, we present the numerical algorithms
that are useful for the study of quantum thermalization. 
We conclude the paper with a summary in Section~\ref{sec:summary}.

\section{Spin-1/2 XXZ model}\label{sec:model}
We consider a one-dimensional array of $L$ spins described by the Pauli
matrices $\{\op\sigma^x_l,\op\sigma^y_l,\op\sigma^z_l\}$ with
$l=0,\cdots,L-1$.
The XXZ Hamiltonian reads as
\begin{equation}\label{HXXZ}
\begin{split}
\opH &= -\frac{J}{2} \sum_{l=0}^{L-1} \left[ 
      \opsigma_l^x \opsigma_{l+1}^x +
      \opsigma_l^y \opsigma_{l+1}^y +
      \Delta \opsigma_l^z \opsigma_{l+1}^z \right] \\
              & = -J \sum_{l=0}^{L-1} \left[
      \opsigma_l^+ \opsigma_{l+1}^- + 
      \opsigma_l^- \opsigma_{l+1}^+ +
      \frac{\Delta}{2} \opsigma_l^z \opsigma_{l+1}^z \right] ,
\end{split}
\end{equation}
where $\opsigma_l^{\pm} \equiv 
(\opsigma_l^x \pm i \opsigma_l^y )/2$ are the raising and the lowering
operators, $J=1$ is an overall coupling strength, 
and $\Delta$ is an anisotropy factor. 
Quantum mechanical operators are marked with the symbol \^{}.
We assume the periodic boundary condition where site $l+L$ is identified as  
site $l$. For simplicity, we consider even $L$ only.

In terms of the eigenvalues of $\op\sigma^z$, each site can be in either an
up or a down state. The spin state can be
interpreted as the occupation state of a boson 
subjected to a hard-core repulsion. That is, identifying $\op\sigma_l^{+} =
\op{b}^\dagger_l$ and $\op\sigma_l^-=\op{b}_l$ with the boson creation and
annihilation operators $\op{b}^\dagger_l$ and $\op{b}_l$, respectively, 
one can rewrite the Hamiltonian as
$$
\op{H} = -\sum_{l=0}^{L-1} \left[ \op{b}_l^\dagger \op{b}_{l+1} + \op{b}_l
\op{b}_{l+1}^\dagger + \frac{\Delta}{2} ( 2 \op{n}_l-1) (2\op{n}_{l+1}-1)
\right]
$$
with the number operator $\op{n}_l = \op{b}_l^\dagger \op{b}_l$. 
They satisfy the commutation relations $[\op{b}_l,\op{b}_m] =
[\op{b}_l^\dagger,\op{b}_m^\dagger]=0$ and $[\op{b}_l^\dagger,\op{b}_l] =
\delta_{lm}$. The hard-core repulsion is implemented by setting 
$(\op{b}_l^\dagger)^2 = (\op{b}_l)^2 = 0$. The Hilbert space dimension is
$D=2^L$.

The XXZ model or the hard-core boson model with  
nearest-neighbor interactions is exactly solvable using the Bethe
ansatz~\cite{Izyumov:1988vi}.
Despite solvability, we explain numerical algorithms for the explicit
example of the XXZ model.
All the numerical techniques explained can be generalized to nonintegrable 
systems easily. 
In Section~\ref{sec:thermal}, we cover the XXZ Hamiltonian 
with next-nearest neighbor interactions which is nonintegrable.

A site $l$ may be in either a $|1\rangle_l$ or a $|0\rangle_l$ state, where
$|1(0)\rangle_l$ denotes the eigenstate of 
$\op{n}_l = (1+\op\sigma_l^z)/2$ with eigenvalue $1~(0)$.
The Hilbert space is spanned by the states
$|n_{L-1}\rangle_{L-1}\otimes |n_{L-2}\rangle_{L-2} \otimes\cdots\otimes|n_0\rangle_0$ or
$|\bm{n}\rangle$ with $\bm{n} = n_{L-1}n_{L-2}\cdots n_0$ in short.
Here, $\bm{n}$ may be regarded as a binary representation of an integer 
ranging from 0 to $2^L-1$.
Thus, a basis state $|\bm{n}\rangle$ is represented by an integer variable,
which is convenient in numerical algorithms.

As a basic task, we explain how to construct the $D \times D$ 
Hamiltonian matrix $\mathsf{H} = \{ H_{\bm{mn}} = \langle
\bm{m}|\op{H}| \bm{n}\rangle\}$ 
in the integer representation.
The $\bm{n}$th column is determined from the relation
$\op{H} |\bm{n}\rangle = \sum_{\bm{m}} H_{\bm{mn}} |\bm{m}\rangle$, i.e., from
the state generated by the action of $\op{H}$ on $|\bm{n}\rangle$. 
The pseudocode in Alg.~\ref{alg:actH} illustrates the method to construct the 
list {\texttt{output}} containing the pairs of basis states and the Hamiltonian 
matrix elements.
The $\Delta$ term is diagonal in the number representation.
The nontrivial part is to find the state $(\op{b}_l^\dagger\op{b}_{l+1} +
\op{b}_l\op{b}^\dagger_{l+1}) | \bm{n}\rangle$. It leads to the null state if
$n_{l} = n_{l+1}$. Otherwise, it leads to a state 
represented by the integer $\bm{m}$, which is obtained by
flipping the $l$th and $(l+1)$th bits of $\bm{n}$.
Applying the algorithm for all $\bm{n}$, one obtains the Hamiltonian matrix in
$O(L\times 2^L)$ operations. 
As a reference, we present the Hamiltonian matrix 
for $L=4$ explicitly in Eq.~\eqref{H4full}. 

If the Hamiltonian includes the next-nearest neighbor interactions, one needs
to modify the part for the diagonal matrix element and to add an additional
bit flipping operation in Alg.~\ref{alg:actH}, which is straightforward.
We stress that the model dependence is encoded only in Alg.~\ref{alg:actH}.
The algorithms in the remaining sections are generic for any systems sharing
the same symmetry property.

\begin{figure}
\begin{algorithm}[H]
\caption{Acting Hamiltonian on $|\bm{n}\rangle$}\label{alg:actH}
\begin{algorithmic}
\Procedure{actingH}{$\bm{n}$}
\State{$\mbox{{\tt output = (state,weight)}} = \{\}$}
\State{$\mbox{diag} =  - (\Delta/2)\sum_{l=0}^{L-1} (2n_l-1)(2n_{l+1}-1)$}
\State{append ($\bm{n}$,diag) to {\tt output}}
\For{($l=0$ to $L-1$)}
   \If {($n_l \neq n_{l+1}$)}
      \State {$\bm{m}  = \mbox{bit\_flip}(\bm{n},l,l+1)$}
      \State{append ($\bm{m},-1$) to {\tt output}}
   \EndIf
\EndFor
\State{Return({\tt output})}
\EndProcedure
\end{algorithmic}
\end{algorithm}
\end{figure}

\section{Block diagonalization}\label{sec:symmetry}
\subsection{Particle Number Conservation}
Conservation and symmetry are useful.
First of all, the number operator $\op{N} = \sum_l \op{b}_l^\dagger\op{b}_l$ 
commutes with $\op{H}$ and its eigenvalue $N=0,\cdots,L$ is a good quantum number. 
Thus, the Hamiltonian matrix $\mathsf{H}$ has a
block-diagonal form 
\begin{equation}
\mathsf{H} = \mathsf{H}_0 \oplus \cdots \oplus \mathsf{H}_N \oplus \cdots
\oplus \mathsf{H}_L, 
\end{equation} 
where $\mathsf{H}_N$ denotes the Hamiltonian matrix 
in the $D_N ={}_L C_N$ dimensional 
subspace spanned by states with $N$ particles. 
The subspace will be called the $N$-particle
sector and be denoted by $\mathcal{S}_N$. 

In order to construct $\mathsf{H}_N$, one needs to construct the basis set 
for $\mathcal{S}_N$. The method for doing this is explained in 
the pseudocode in Alg.~\ref{alg:basisN}, where only $N$-particles states 
among all $|\bm{n}\rangle$'s are stored in the list $\texttt{basisN}$. 
Using the list, one can easily reconstruct the block Hamiltonian matrix 
$\mathsf{H}_N$ by following Alg.~\ref{alg:HN}. 
The block-diagonal form of $\mathsf{H}$ for $L=4$ is presented in Eq.~\eqref{H4N}.

\begin{figure}
\begin{algorithm}[H]
\caption{Constructing the basis set for $\mathcal{S}_N$}\label{alg:basisN}
\begin{algorithmic}
\Procedure{build-basisN}{$L,N$}
 \State{{\tt basisN} = \{\}}
 \For{($\bm{n} = 0 \mbox{ to } 2^L -1$)}
   \If {$(\sum_l n_l = N)$}
      \State{append $\bm{n}$ to {\tt basisN}} 
   \EndIf
 \EndFor
 \State{Return({\tt basisN})}
\EndProcedure
\end{algorithmic}
\end{algorithm}
\end{figure}

\begin{figure}
\begin{algorithm}[H]
\caption{Block Hamiltonian $\mathsf{H}_N$}\label{alg:HN}
\begin{algorithmic}
\Procedure{buildHN}{$L,N$}
 \State{$\mbox{HN} = \{0\}$}
 \State{{\tt basisN} $\gets$ {\textsc{build-basisN($L,N$)}}}

\For{($\bm{n}\in {\tt basisN}$)}
 \State{$b = \mbox{find\_index\_in\_basisN}(\bm{n})$}
 \State{{\tt output} $\gets {\textsc{actingH}}(\bm{n})$}
  \For{$((\bm{m},h) \in {\tt output})$}
     \State{$a = \mbox{find\_index\_in\_basisN}(\bm{m})$}
     \State{$\mbox{HN}[a,b] = \mbox{HN}[a,b] + h$}
  \EndFor
\EndFor
\State{Return(HN)}
\EndProcedure
\end{algorithmic}
\end{algorithm}
\end{figure}

\subsection{Translational Symmetry}
Translational symmetry means invariance under the
shift operator $\op{T}$ defined as
\begin{equation}
\op{T} |\bm{n}\rangle = |T(\bm{n})\rangle ,
\end{equation}
where the function $\bm{m} = T(\bm{n})$ shifts binary bits of $0\leq \bm{n}
<2^L$ by a unit distance~($m_l = n_{l-1}$). 
Because $\op{T}^L=1$, its 
eigenvalue takes the values of
\begin{equation}
\omega_k = \exp\left( \frac{2\pi i k}{L} \right)  \quad (k=0,1,\cdots,L-1) ,
\end{equation}
where $k$ is called the wave number.
The three operators $\op{H}$, $\op{N}$, and $\op{T}$ commute with
one another. 
Thus, if one chooses the simultaneous eigenstates of $\op{N}$ and
$\op{T}$ as the basis set, then one can block-diagonalize 
$\mathsf{H}_N$ to the form 
\begin{equation}
\mathsf{H}_N = \mathsf{H}_{N,0}\oplus\cdots\oplus\mathsf{H}_{N,k}
           \oplus\cdots\oplus\mathsf{H}_{N,L-1} ,
\end{equation}
where $\mathsf{H}_{N,k}$ is the Hamiltonian matrix in the subspace
$\mathcal{S}_{N,k}$ characterized by the particle number $N$ and the wave
number $k$. 

In order to construct the basis set for $\mathcal{S}_{N,k}$,
we define the equivalence relation:
$|\bm{n}\rangle$ and $|\bm{m}\rangle$ are equivalent under translation 
if $|\bm{m}\rangle = \op{T}^l |\bm{n}\rangle$ or $\bm{m} = T^l (\bm{n})$ 
for an integer $l$.
With equivalence, the basis states of $\mathcal{S}_N$ can be grouped into 
distinct sets of mutually equivalent states.
Such a set is called the equivalent class~(EC) and has the form
$\{|\bm{n}\rangle,|T(\bm{n})\rangle, \cdots,|T^l(\bm{n})\rangle, \cdots\}$.
An EC is represented with the state $|\bar{\bm{n}}\rangle$ 
where $\bar{\bm{n}} = \min_l[ T^l(\bm{n})]$, which is called 
the representative state~(RS). 
The size of an EC is denoted by $p(\bar{\bm{n}})$, which is called
the period because $T^{p(\bar{\bm{n}})}(\bar{\bm{n}}) = \bar{\bm{n}}$.
The representative states for $L=4$ are listed in Table~\ref{RSL4}.

\begin{table}
\caption{Representative states $\bar{\bm{n}}$ for $L=4$
and the members of the EC
represented by $\bar{\bm{n}}$ and the period.}\label{RSL4}
\begin{ruledtabular}
\begin{tabular}{c c c c}
$N$ & $\bar{\bm{n}}$ & $\bm{n}$ & period $p(\bar{\bm{n}})$ \\ \hline
0   & 0000  & 0000 & 1 \\ \hline
\multirow{4}{*}{1} & \multirow{4}{*}{0001}  & 0001 & \multirow{4}{*}4 \\ 
    &       & 0010 &   \\
    &       & 0100 &   \\
    &       & 1000 &   \\ \hline
\multirow{6}{*}{2} & \multirow{4}{*}{0011} & 0011 & \multirow{4}{*}4 \\
                   &                       & 0110 &  \\
                   &                       & 1100 &  \\
                   &                       & 1001 &  \\ \cline{2-4}
                   & \multirow{2}{*}{0101} & 0101 & \multirow{2}{*}2 \\
                   &                       & 1010 &  \\ \hline
\multirow{4}{*}{3} & \multirow{4}{*}{0111}  & 0111 & \multirow{4}{*}4 \\ 
    &       & 1110 &  \\
    &       & 1101 &  \\
    &       & 1011 &  \\ \hline
4   & 1111  & 1111 & 1 
\end{tabular}
\end{ruledtabular}
\end{table}

The simultaneous eigenstates of $\op{N}$ and $\op{T}$ with eigenvalues $N$ and
$\omega_k$ constitute the basis set of $\mathcal{S}_{N,k}$. They are given
by
\begin{equation}\label{basis_Nk}
|\bar{\bm{n}},k\rangle = Y(\bar{\bm{n}}) \sum_{l=0}^{L-1} \omega_k^{-l} ~
\op{T}^l |\bar{\bm{n}}\rangle ,
\end{equation}
where $|\bar{\bm{n}}\rangle$ is a RS in $\mathcal{S}_N$ and 
\begin{equation}\label{nomal_c}
Y(\bar{\bm{n}}) = \frac{\sqrt{p(\bar{\bm{n}})}}{L} .
\end{equation}
We should note that only the RS's satisfying the commensurability condition
\begin{equation}
k p(\bar{\bm{n}}) = \mbox{(integer)} \times L 
\end{equation}
yield the basis states. If the commensurability condition does not hold, 
the state in Eq.~\eqref{basis_Nk} becomes a null state.
With Eq.~\eqref{basis_Nk}, storing the list of RS's for the
basis set for $\mathcal{S}_{N,k}$ is sufficient.
Algorithm~\ref{alg:basisNk} shows the pseudocode to generate the list
\texttt{basisNk}.
 
\begin{figure}
\begin{algorithm}[H]
\caption{List of representative states contributing to the basis set of 
$\mathcal{S}_{N,k}$}\label{alg:basisNk}
\begin{algorithmic}
\Procedure{build-basisNk}{$L,N,k$}
\State{{\tt basisNk} = \{\}}
\State{{\tt basisN} $\gets$ \textsc{build-basisN}$(L,N)$}
\For{($\bm{n} \in \mbox{\tt basisN}$)}
 \State{$\bm{m} = \min_{l=0}^{L-1} T^l(\bm{n})$ : {\em RS of $\bm{n}$}}
 \If{(($\bm{n} = \bm{m}$) and (commensurability))} 
   \State{append $\bm{n}$ to {\tt basisNk}} 
 \EndIf
\EndFor
\State{Return({\tt basisNk})}
\EndProcedure
\end{algorithmic}
\end{algorithm}
\end{figure}

The Hamiltonian matrix elements 
\begin{equation}
H^{(N,k)}_{\bar{\bm{m}}\bar{\bm{n}}} = \langle \bar{\bm{m}},k| \op{H} |
\bar{\bm{n}},k\rangle
\end{equation}
are also easily read off from the outcome states of the product
$\op{H}|\bar{\bm{n}},k\rangle$. Applying $\op{H}$ on $|\bar{\bm{n}},k\rangle$
in Eq.~\eqref{basis_Nk}, we obtain
\begin{equation}\label{Hkprod}
\begin{aligned}
\op{H}|\bar{\bm{n}},k\rangle & = Y(\bar{\bm{n}}) \left[\sum_{l=0}^{L-1}
\omega_k^{-l} \op{T}^l \right] \op{H} | \bar{\bm{n}}\rangle \\
  &= Y(\bar{\bm{n}}) \sum_{\bm{m}} \left[\sum_{l=0}^{L-1} \omega_k^{- l}
\op{T}^l
\right] H_{\bm{m}\bar{\bm{n}}} |\bm{m}\rangle ,
\end{aligned}
\end{equation}
where we have used translational symmetry in the first line.
Each microscopic state $|\bm{m}\rangle$ is written as
$|\bm{m}\rangle = \op{T}^{d(\bm{m})} 
|\bar{\bm{m}}\rangle$, where $|\bar{\bm{m}}\rangle$ is the RS of $\bm{m}$ and 
$d(\bm{m})$ is the distance of $\bm{m}$ from $\bar{\bm{m}}$.
Then, Eq.~\eqref{Hkprod} becomes
\begin{equation}
\begin{aligned}
\op{H}|\bar{\bm{n}},k\rangle &= Y(\bar{\bm{n}}) \sum_{\bm{m}}
            \left[\sum_{l=0}^{L-1} \omega_k^{- l} \op{T}^{l+d(\bm{m})}
             \right] H_{\bm{m}\bar{\bm{n}}} |\bar{\bm{m}}\rangle \\
    &= Y(\bar{\bm{n}}) \sum_{\bm{m}} \omega_k^{d(\bm{m})}
            H_{\bm{m}\bar{\bm{n}}}
            \left[\sum_{l=0}^{L-1} \omega_k^{- l} \op{T}^l
             |\bar{\bm{m}}\rangle\right] \\
 &= \frac{Y(\bar{\bm{n}})}{Y(\bar{\bm{m}})} \sum_{\bm{m}}
\omega_k^{d(\bm{m})} H_{\bm{m}\bar{\bm{n}}} | \bar{\bm{m}},k\rangle .
\end{aligned}
\end{equation}
Thus, the matrix elements are given by
\begin{equation}
H^{(N,k)}_{\bar{\bm{m}}\bar{\bm{n}}} = \frac{Y(\bar{\bm{n}})}{Y(\bar{\bm{m}})}
\sum_{\bm{m}}' \omega_k^{d(\bm{m})}~ H_{\bm{m}\bar{\bm{n}}} ,
\end{equation}
where the primed summation is over all states $|\bm{m}\rangle$ belonging to
the EC represented by RS $|\bar{\bm{m}}\rangle$.
The pseudocode in Alg.~\ref{alg:HNk} explains how the matrix $\mathsf{H}_{N,k}$ 
is constructed. As a reference, we present the block
diagonal form of $\mathsf{H}_{N=2}$ for $L=4$ in Eq.~\eqref{H4N2k}.

\begin{figure}
\begin{algorithm}[H]
\caption{Matrix elements of $\mathsf{H}_{N,k}$}\label{alg:HNk}
\begin{algorithmic}
\Procedure{buildHNk}{{$L,N,k$}}
\State{HNk = \{0\}}
\State{{\tt basisNk} $\gets$ \textsc{build-basisNk($L,N,k$)}}
\For{($\bar{\bm{n}} \in{\tt basisNk}$)}
 \State{$b = \mbox{find\_index\_in\_basisNk}(\bar{\bm{n}})$ }
   \State{{\tt output} $\gets$ \textsc{actingH}($\bar{\bm{n}}$)}
   \For{($(\bm{m},h) \in {\tt output}$)}
      \State{$\bar{\bm{m}} = \mbox{RS of $\bm{m}$}$}
      \State{$d$ = distance from $\bar{\bm{m}}$ to $\bm{m}$}
      \If{$(\bar{\bm{m}} \in \mbox{\tt basisNk})$}
        \State{$a = \mbox{find\_index\_in\_basisNk($\bar{\bm{m}}$)}$}
        \State{{HNk}[a,b] = \mbox{HNk}[a,b] + 
               $\frac{Y(\bar{\bm{n}})}{Y(\bar{\bm{m}})} \omega_k^d h$}
      \EndIf
   \EndFor
\EndFor
\State{Return(HNk)}
\EndProcedure
\end{algorithmic}
\end{algorithm}
\end{figure}

We list the dimensionality of the symmetry sectors in
Table~\ref{dimension}. Among all particle number sectors $\mathcal{S}_N$, 
the half-filling sector ($N=L/2$) is the largest.
Within the half-filling sector, the translationally invariant sector 
$\mathcal{S}_{L/2,0}$ is the largest. Roughly
speaking, the dimensionality scales as $|\mathcal{S}_{N=L/2}| =
O(L^{-1/2} 2^L)$ and $|\mathcal{S}_{N=L/2,k=0}| = O(L^{-3/2} 2^L)$. 
Particle number conservation and translational symmetry reduce 
the Hilbert space dimensionality by the factor $O(L^{3/2})$.
 
\begin{table}
\caption{Hilbert space dimensionality}\label{dimension}
\begin{ruledtabular}
\begin{tabular}{c|rrrr}
$L$ & $|\mathcal{S}|$ & $|\mathcal{S}_{N=L/2}|$ &
$|\mathcal{S}_{N=L/2,k=0}|$ & $|\mathcal{S}_{L/2,0,+1,+1}|$ \\ \hline
4  & 16 & 6 & 2 & 2 \\
6  & 64 & 20 & 4 & 3 \\
8  & 256 & 70 & 10 & 7 \\
10 & 1,024 & 252 & 26 & 13 \\
12 & 4,096 & 924 & 80 & 35 \\
14 & 16,384 & 3,432 & 246 & 85 \\
16 & 65,536 & 12,870 & 810 & 257 \\
18 & 262,144 & 48,620 & 2,704 & 765 \\
20 & 1,048,576 & 184,756 & 9,252 & 2,518 \\
22 & 4,194,304 & 705,432 & 32,066 & 8,359 \\
24 & 16,777,216 & 2,704,156 & 112,720 & 28,968  \\
26 & 67,108,864  & 10,400,600 & 400,024 & 101,340 \\
\end{tabular}
\end{ruledtabular}
\end{table}

\subsection{Inversion and Reflection Symmetry}
We can make use of the additional discrete symmetry.
The system has spin reversal symmetry or, equivalently, 
particle-hole symmetry.
That is, the Hamiltonian commutes with the spin reversal operator:
\begin{equation}
\op{X} = \op{X}^\dagger = 
\prod_{l=0}^{L-1} \op\sigma_l^x = \prod_{l=0}^{L-1} (\op{b}_l^\dagger +
\op{b}_l) .
\end{equation}
The system is also symmetric under spatial reflection, and the
Hamiltonian commutes with the reflection operator $\op{R} = \op{R}^\dagger$ defined as
\begin{equation}
\op{R} \op{O}_l \op{R} = \op{O}_{L-1-l} 
\end{equation}
for any local operators $\op{O}_l$ at site $l$.
Because $\op{X}^2 = \op{R}^2 = 1$, their eigenvalues are
$X=\pm 1$ for $\op{X}$ and $R=\pm 1$ for $\op{R}$.

\begin{figure}
\includegraphics*[width=0.4\columnwidth]{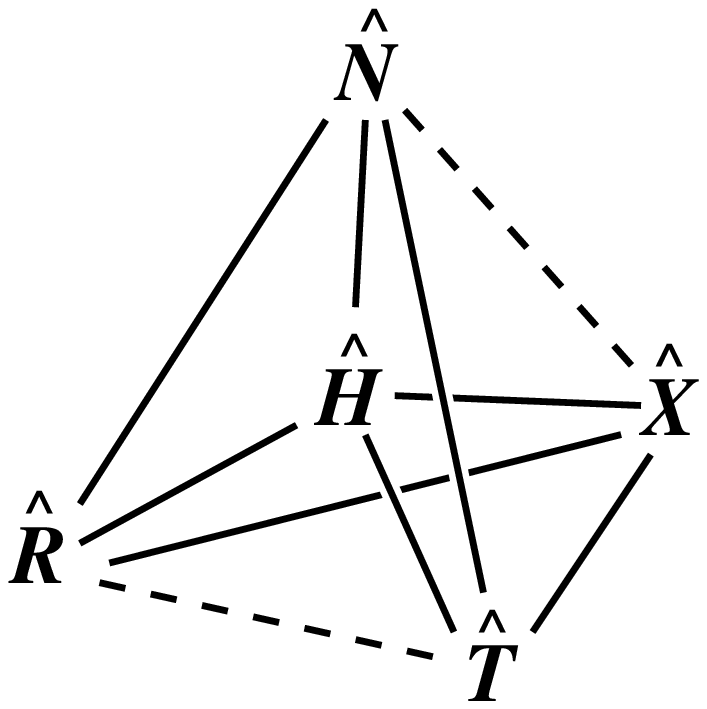}
\caption{Commutation relations among the Hamiltonian and the symmetry
operators. Commuting operators are connected with solid lines while
noncommuting operators are connected with dashed lines.}\label{commutation}
\end{figure}

Unfortunately, these discrete symmetry operators do not commute with all the other
symmetry operators: $[\op{R},\op{T}]\neq 0$ and $[\op{N},\op{X}]\neq 0$.
Instead, they satisfy the relations
\begin{equation}\label{symm_rel}
\begin{aligned}
\op{X}^\dagger \left(\op{N}-L/2\right) \op{X} &= -\left(\op{N}-{L}/2\right)  , \\
\op{R}^\dagger \op{T} \op{R} &= \op{T}^{-1} .
\end{aligned}
\end{equation}
Figure~\ref{commutation} summarizes the commutation properties. 

\begin{figure}
\includegraphics[width=0.9\columnwidth]{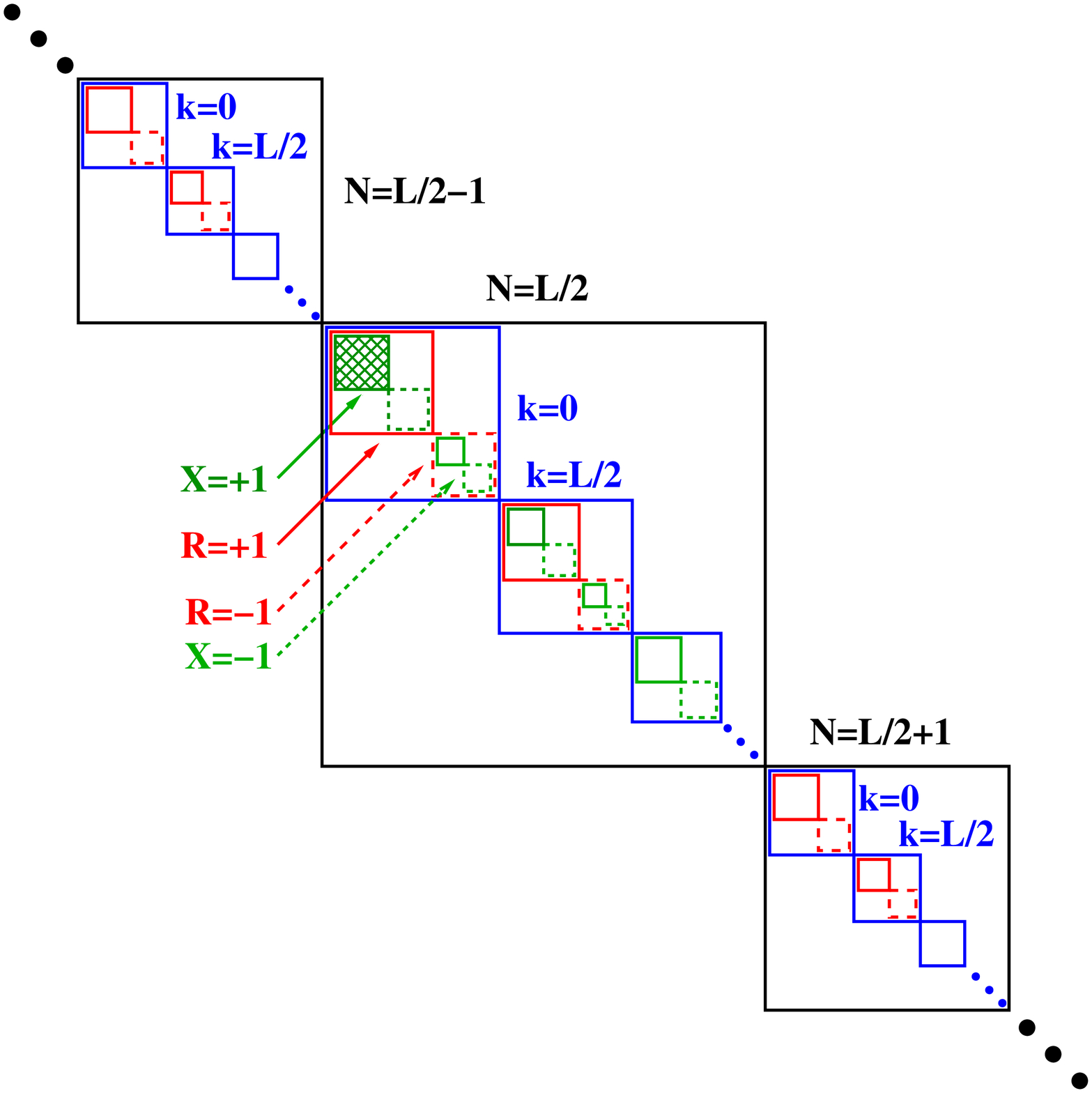}
\caption{Block diagonal structure of the Hamiltonian matrix. The shaded area
denotes the maximum symmetry sector.}\label{block}
\end{figure}

Despite the nontrivial commutation property, discrete symmetry is still
useful.  From Eq.~\eqref{symm_rel},
one finds that $\op{N}$ and $\op{X}$ commute within the {\em half-filling
sector} with $N=L/2$. 
Thus, inside the half-filling sector, $(\op{H},\op{N},\op{T},\op{X})$ are 
mutually commuting, and the Hamiltonian can be block-diagonalized as
\begin{equation}
\mathsf{H}_{N=L/2,k} = \mathsf{H}_{N=L/2,k,X=+1} \oplus \mathsf{H}_{N=L/2,k,X=-1}
\end{equation}
for any $k$. 
One also finds that $\op{R}$ and $\op{T}$ commute within 
the sectors $\mathcal{S}_{N,k=0}$ and $\mathcal{S}_{N,k=L/2}$ 
where $\op{T} = \op{T}^{-1}$. Thus, the Hamiltonian in the symmetric~($k=0$) and 
the anti-symmetric~($k=L/2$) sectors under translation can be decomposed as
\begin{equation}
\mathsf{H}_{N,k} = 
 \mathsf{H}_{N,k,R=+1} \oplus \mathsf{H}_{N,k,R=-1}
\end{equation}
for any $N$. 
Especially, within the subspace with $(N,k) = (L/2,0)$ or $(L/2,L/2)$, 
all the symmetry operators mutually commute with one another. 
The block-diagonal structure of $\mathsf{H}$ is illustrated in
Fig.~\ref{block}.

\subsection{Maximum Symmetry Sector}
We focus on the subspace $\mathcal{S}_{N=L/2,k=0,R=+1,X=+1}$, which will be called 
the maximum symmetry sector~(MSS). This sector corresponds to the shaded block in
Fig.~\ref{block}. 
In order to incorporate all the symmetry, we extend the concept of the equivalent class. 
Suppose that $|\bar{\bm{n}}\rangle$ is a RS in the half-filling sector. 
The symmetry operations $\op{X}$, $\op{R}$, and $\op{R}\op{X}$ map
$|\bar{\bm{n}}\rangle$ to a member of other ECs represented by 
$|\bar{\bm{n}}_X\rangle$, $|\bar{\bm{n}}_R\rangle$, and 
$|\bar{\bm{n}}_{RX}\rangle$, respectively.
All the involved ECs merge into a single set, which is defined as the super 
equivalent class~(SEC). A SEC is represented by the super representative
state~(SRS) $|\tilde{\bm{n}}\rangle$, where
\begin{equation}
\tilde{\bm{n}} = \min\left[ \bar{\bm{n}}, \bar{\bm{n}}_R, \bar{\bm{n}}_X,
\bar{\bm{n}}_{RX}\right] .
\end{equation}
A degeneracy may exist among the four numbers $\bar{\bm{n}}, \bar{\bm{n}}_R,
\bar{\bm{n}}_X$, and $\bar{\bm{n}}_{RX}$.
The number of distinct elements among the four will be
denoted as the multiplicity factor $q(\tilde{\bm{n}})$ of the SEC. 
Then, the states
\begin{equation}\label{srs}
|\tilde{\bm{n}}\rangle_{\rm MSS} = Z(\tilde{\bm{n}}) (1+\op{X})(1+ \op{R})
    \left(\sum_{l=0}^{L-1}\op{T}^l\right) |\tilde{\bm{n}}\rangle
\end{equation}
for all SRS's form the basis set for the MSS. 
The normalization factor is given by
\begin{equation}\label{nomal_b}
Z(\tilde{\bm{n}}) = \frac{\sqrt{q(\tilde{\bm{n}})}}{4} {Y(\tilde{\bm{n}})} 
= \frac{\sqrt{p(\tilde{\bm{n}}) q(\tilde{\bm{n}})}}{4 L}
\end{equation}
with the function $Y$ in Eq.~\eqref{nomal_c}.
The pseudocode to find the list of SRS's is presented in Alg.~\ref{alg:srs}.
In Table~\ref{srsL8}, we list the SRS's in the MSS for $L=8$, along with the
normalization constants.
The dimensionality of the MSS is listed in Table~\ref{dimension}.

\begin{figure}
\begin{algorithm}[H]
\caption{List of SRS's for the basis set of the MSS}
\label{alg:srs}
\begin{algorithmic}
\Procedure{build-SRSlistMSS}{$L$}

\State{{\tt SRSlistMSS} = \{\}}
\State{$\mbox{\tt basisNk} \gets$ \textsc{build-basisNk($L,N=L/2,k=0$)}}
\For{($\bar{\bm{n}} \in \mbox{\tt basisNk}$)}
 \If{$(\bar{\bm{n}} \leq
     \min[\bar{\bm{n}}_X,\bar{\bm{n}}_R,\bar{\bm{n}}_{RX}])$}
     \State{append $\bar{\bm{n}}$ to {\tt SRSlistMSS}}
 \EndIf
\EndFor
\State{Return({\tt SRSlistMSS})}
\EndProcedure
\end{algorithmic}
\end{algorithm}
\end{figure}

\begin{table}
\caption{Super representative states in the MSS for $L=8$}\label{srsL8}
\begin{ruledtabular}
\begin{tabular}{cccc}
SRS & period $p$ & multiplicity $q$ & normalization $Z$ \\ \hline
$|00001111\rangle$ & 8 & 1 & $\sqrt{2}/16$  \\
$|00010111\rangle$ & 8 & 2 & $1/8$ \\
$|00011011\rangle$ & 8 & 2 & $1/8$ \\
$|00101101\rangle$ & 8 & 1 & $\sqrt{2}/16$  \\
$|00110011\rangle$ & 4 & 1 & $1/16$  \\
$|00101011\rangle$ & 8 & 2 & $1/8$  \\
$|01010101\rangle$ & 2 & 1 & $\sqrt{2}/32$ 
\end{tabular}
\end{ruledtabular}
\end{table}

Applying the Hamiltonian to a basis state, one obtains
$$
\op{H} | \tilde{\bm{n}}\rangle_{\rm MSS} = Z(\tilde{\bm{n}}) \sum_{\bm{m}}
H_{\bm{m},\tilde{\bm{n}}}
(1+\op{X})(1+\op{R})\left(\sum_{l=0}^{L-1}\op{T}^l\right) |\bm{m}\rangle .
$$
Thus, the Hamiltonian matrix elements in the MSS are given by
\begin{equation}
H^{({\rm MSS})}_{\tilde{\bm{m}}\tilde{\bm{n}}} =
\frac{Z(\tilde{\bm{n}})}{Z(\tilde{\bm{m}})} \sum'_{\bm{m}}
H_{\bm{m}\tilde{\bm{n}}} ,
\end{equation}
where the primed summation is over all states $|\bm{m}\rangle$
belonging to the same SEC as $|\tilde{\bm{m}}\rangle$. 
In Alg.\ref{alg:hmss}, we present a pseudocode to construct 
the matrix $\mathsf{H}_{\rm MSS}$. 
The explicit expression  of $\mathsf{H}_{\rm MSS}$ for $L=8$ is shown in
Eq.~(\ref{HmssL8}).
\begin{figure}
\begin{algorithm}[H]
\caption{Matrix elements for $\mathsf{H}_{MSS}$}\label{alg:hmss}
\begin{algorithmic}
\Procedure{build-Hmms}{{$L$}}
 \State{{\tt SRSlistMSS} $\gets$ \textsc{build-SRSlistMSS}($L$)}

 \For{($\tilde{\bm{n}} \in {\tt SRSlistMSS}$)}
  \State{$b = \mbox{find\_index\_in\_SRSlistMSS}(\tilde{\bm{n}})$}
  \State{$\mbox{\tt output} \gets$
         \textsc{ActingH($\tilde{\bm{n}}$)}}
  \For{($(\bm{m},h) \in {\tt output}$)}
   \State{$\tilde{\bm{m}} = \mbox{SRS for }\bm{m}$}
   \State{$a = \mbox{find\_index\_in\_SRSlistMSS}(\tilde{\bm{m}})$}
   \State{HMMS[a,b] = HMMS[a,b] + $h Z(\tilde{\bm{n}})/Z(\tilde{\bm{m}})$}
  \EndFor
 \EndFor
\EndProcedure
\end{algorithmic}
\end{algorithm}
\end{figure}

The basis set in the other symmetry sectors can be constructed in a similar way. 
For each SRS $|\tilde{\bm{n}}\rangle$, one can consider a state
\begin{equation}
|\tilde{\bm{n}}\rangle_{w,X,R} \propto
(1+ X \op{X}) (1+R\op{R}) \left( \sum_{l=0}^{L-1} (w \op{T})^l \right)
|\tilde{\bm{n}}\rangle
\end{equation}
with $X=\pm 1$, $R = \pm 1$, and $\omega = \pm 1$. It may lead to a null state
or a nonvanishing state depending on $\tilde{\bm{n}}$ and $(X,R,\omega)$.
The set of nonvanishing states form the basis set for
$\mathcal{S}_{N=L/2,k=0,R=\pm 1,X=\pm 1}$ or ${S}_{N=L/2,k=L/2,R=\pm 1,X=\pm 1}$.
The Hamiltonian matrix can also be constructed similarly, which is not
shown in this paper.

\section{Numerical diagonalization}\label{sec:nd}
We have diagonalized the Hamiltonian matrix constructed in  the way 
explained in the previous section to solve the eigenvalue problem
\begin{equation}
\op{H} | \alpha \rangle = E_\alpha | \alpha \rangle .
\end{equation}
The numerical solution is found by diagonalizing the Hamiltonian matrix 
with the help of computational libraries. 
A computational library takes the Hamiltonian matrix $\mathsf{H}=\{H_{mn}\}$
as an input and then outputs the set of eigenvalues $\{E_\alpha\}$ and 
the unitary matrix $\mathsf{S} = \{S_{n\alpha}\}$, 
where $S_{n\alpha} = \langle n|\alpha\rangle$ 
is the $n$th component of the normalized $\alpha$th eigenstate: 
\begin{equation}\label{eigen_eq}
\sum_n H_{mn} S_{n\alpha} = E_\alpha S_{m\alpha}
\end{equation}
or
\begin{equation}\label{simil_S}
\mathsf{S}^\dagger \mathsf{H}\mathsf{S} = {\rm diag}\{E_\alpha\} 
\end{equation}
in matrix form.

\begin{figure}
\includegraphics*[width=\columnwidth]{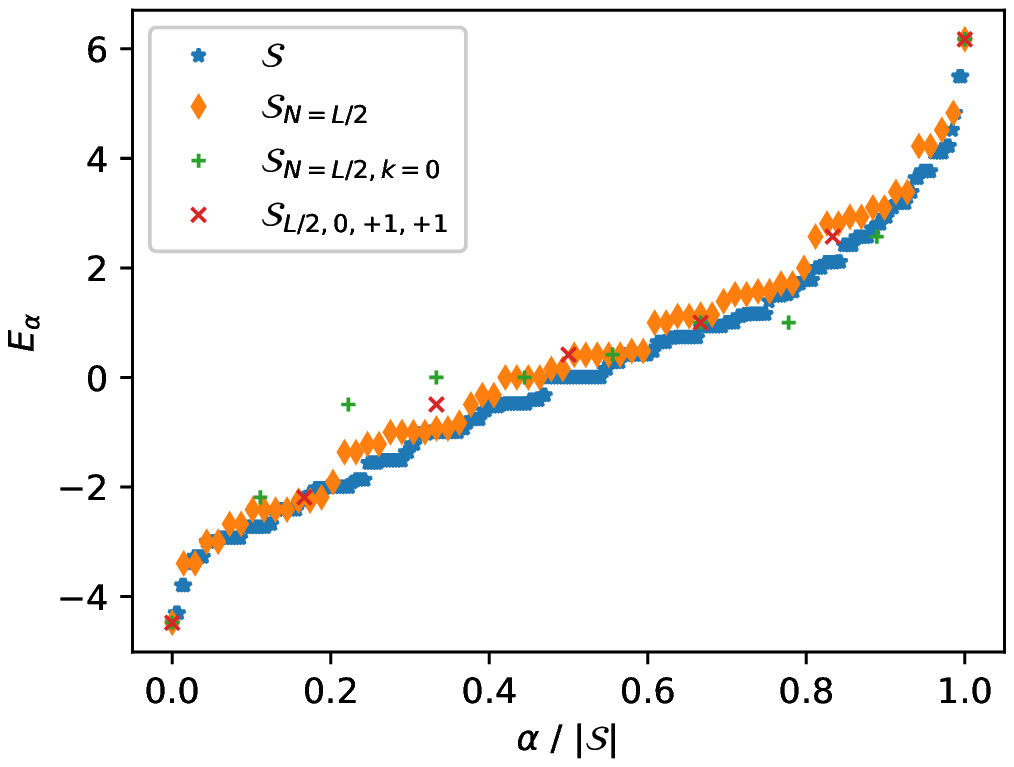}
\caption{Energy eigenvalues are plotted as functions of the energy quantum
number normalized to the Hilbert space dimension. The lattice size is $L=8$.
}\label{E_L8}
\end{figure}

\begin{figure}
\includegraphics*[width=\columnwidth]{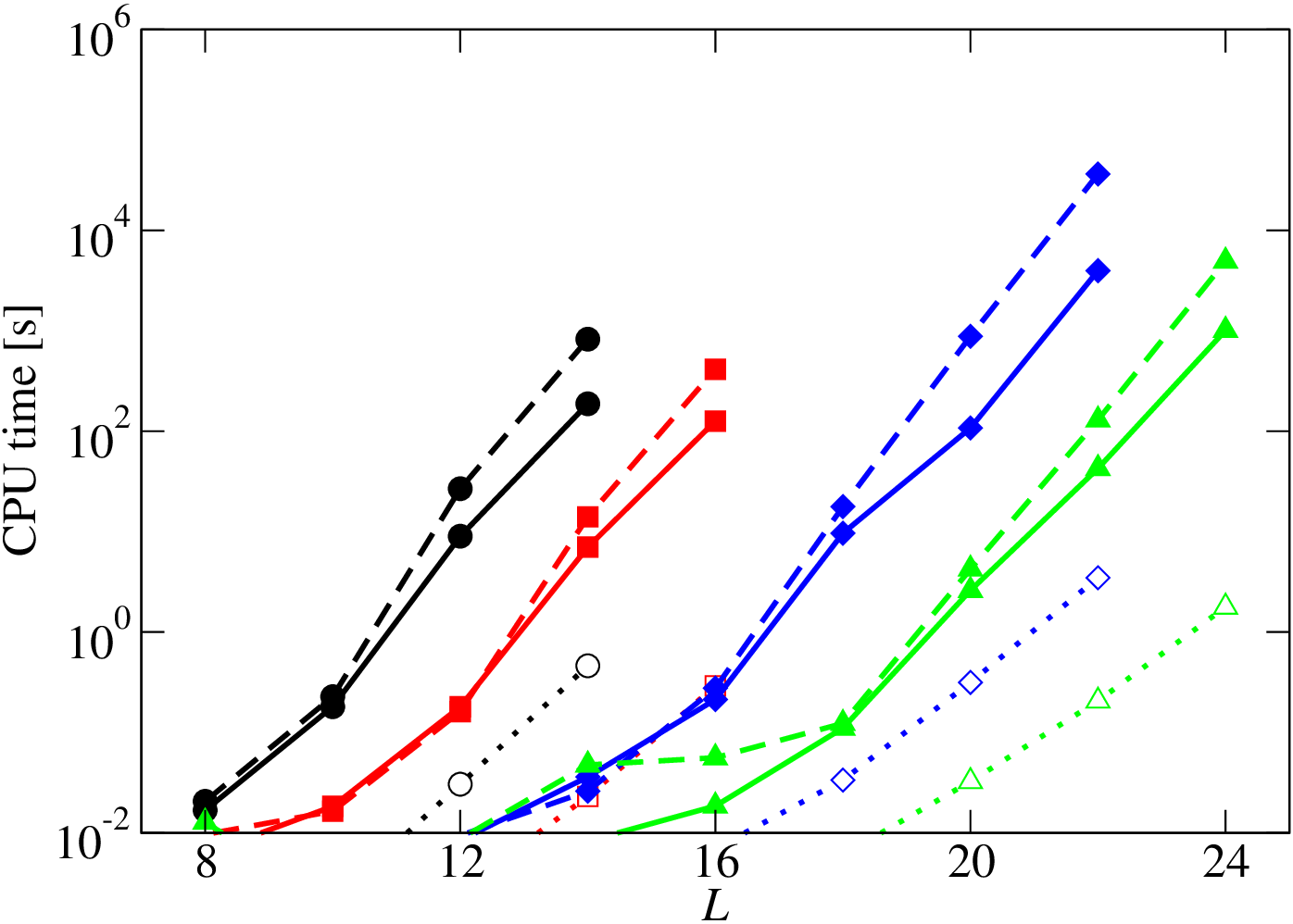}
\caption{CPU times in seconds for matrix construction~(open symbols with dotted
lines), diagonalization without eigenvectors~(filled symbols with solid
lines), and diagonalization with eigenvectors~(filled symbols with dashed
lines). Data for the full Hamiltonian $\mathsf{H}$
and the block Hamiltonians $\mathsf{H}_{N=L/2}$, $\mathsf{H}_{N=L/2,k=0}$, and
$\mathsf{H}_{\rm MMS}$ are marked with circular, square, diamond, and triangular
symbols, respectively.}
\label{fig:cpu_time}
\end{figure}

We have diagonalized the full Hamiltonian matrix $\mathsf{H}$ and the block
Hamiltonians $\mathsf{H}_{N=L/2}$, $\mathsf{H}_{N=L/2,k=0}$, and
$\mathsf{H}_{\rm MSS}$ for the XXZ Hamiltonian in Eq.~\eqref{HXXZ} with $J=1$
and $\Delta = 1/2$. 
The source codes are composed in C language
with the Intel${}^\circledR$ Math Kernel Library~\footnote{The library can be
found in {\texttt https://software.intel.com/en-us/mkl}}. 
The program was run on an Intel${}^\circledR$ Core\texttrademark i9-9900K
processor. With 64-GB memory, the maximum system size 
accessible is $L=24$ for the MSS.
Figure~\ref{E_L8} presents the energy eigenvalue spectrum of the full
Hamiltonian $\mathsf{H}$ for $L=8$ sites 
and of the block Hamiltonians $\mathsf{H}_{N=4}$, 
$\mathsf{H}_{N=4,k=0}$, and $\mathsf{H}_{\rm MMS}$. We have also measured
the CPU times for the matrix construction and the diagonalization with and 
without eigenvectors. The CPU times are plotted in Fig.~\ref{fig:cpu_time}.
Matrix diagonalization uses most of the CPU times. 
Roughly speaking, the CPU time scales algebraically as $D^z$, with the total 
Hilbert space dimension $D=2^L$ with $z\simeq 1.5$ for matrix constructions
and $z\simeq 2.7$ and $2.2$ for diagonalization with and without eigenvectors,
respectively.

\section{Numerical Study of Quantum Thermalization}\label{sec:thermal}
As an application of the numerical technique, we investigate the 
quantum thermalization of the XXZ model with nearest and next-nearest neighbor
interactions.
As mentioned in Section~\ref{sec:intro}, 
quantum systems which thermalize are
believed to obey the ETH, which assumes that
matrix elements $O_{\alpha\gamma} = \langle \alpha |\op{O} | \gamma \rangle$
of a local observable $\op{O}$ in the Hamiltonian eigenstates basis
take the form~\cite{Srednicki:1996kn,Srednicki:1999bo}
\begin{equation}\label{ETH_ansatz}
O_{\alpha\gamma} = O(E_{\alpha\gamma}) \delta_{\alpha\gamma}
+ e^{-S(E_{\alpha\gamma})/2k_B} 
f_O(E_{\alpha\gamma},\omega_{\alpha\gamma}) R_{\alpha\gamma} , 
\end{equation}
where $S$ is the thermodynamic entropy, $O$ and $f_O$ are smooth functions of 
$E_{\alpha\gamma} = (E_\alpha+E_\gamma)/2$ and $\omega_{\alpha\gamma} =
(E_\alpha-E_\gamma)/\hbar$, and $R_{\alpha\gamma}$ are random matrix elements.
The Boltzmann constant $k_B$ and the Planck constant $\hbar$
will be set to unity.
The ETH guarantees that the quantum mechanical expectation value 
is equal to the microcanonical ensemble average. 
The ETH is believed to hold for generic nonintegrable quantum 
systems~\cite{DAlessio:2016gr}. 

We investigate the XXZ spin chain Hamiltonian with nearest and
next-nearest neighbor interactions.
The XXZ Hamiltonian in Eq.~(\ref{HXXZ}) with only nearest neighbor interactions
is a representative example of a nonthermal integrable system~\cite{Rigol:2009ew}. 
It can be made nonintegrable by adding the next-nearest neighbor
couplings~\cite{Yoshizawa:2018js}. 
We consider the Hamiltonian
\begin{equation}\label{Hnnn}
\op{H} = \frac{1}{1+\lambda} \left( \op{H}_{nn} + \lambda \op{H}_{nnn}
\right) ,
\end{equation}
where $\op{H}_{nn}$ is equal to the Hamiltonian in Eq.~\eqref{HXXZ} and
$\op{H}_{nnn}$ denotes the same type of Hamiltonian whose interaction and hopping
ranges are modified to the next-nearest neighbors. With nonzero
$\lambda$ and $\Delta$, the Hamiltonian is known to obey the
ETH~\cite{Yoshizawa:2018js}. 

Even in the presence of the next-nearest neighbor interactions, one can
use the same method to construct the basis set and the Hamiltonian matrix.
One needs to modify Alg.~\ref{alg:actH} in order to include the additional
interaction terms only.  In numerical
calculations hereafter, we fix the values of $J$ to $1$, $\lambda$ to $1$, and $\Delta$ to $1/2$.
In general, one should be able to handle any subspace of the whole Hilbert
space. If the subspace is intractable in a brute-force way, one needs to
decompose the subspace into symmetry sectors, as explained in
Section~\ref{sec:symmetry}. The MSS has the largest dimensionality among all
the symmetry sectors and is the hardest obstacle in numerical studies.
Thus, we mainly focus on the MSS in the numerical demonstration.

\subsection{Energy Eigenstate Expectation Value}

The ETH suggests that the energy eigenstate
expectation value of a local observable $\op{O}$ should depend only on the energy
eigenvalue in the thermodynamic limit.  
We test the hypothesis for two observables: 
the zero momentum distribution function
\begin{equation}\label{mom0}
\op{A} = \frac{1}{L} \sum_{l,m=0}^{L-1} \op{b}_l^\dagger \op{b}_m = 
 \frac{1}{L}\sum_{l,m} \op\sigma_l^+ \op\sigma_m^- 
\end{equation}
and the nearest neighbor interaction energy density
\begin{equation}\label{zz}
\op{B} = \frac{1}{L} \sum_l \op{n} \op{n}_{l+1} =
\frac{1}{L} \sum_l \frac{\op\sigma_l^z+1}{2}\frac{\op\sigma_{l+1}^z+1}{2} .
\end{equation}
These operators commute with the symmetry operators $\op{N}$,
$\op{T}$, $\op{R}$, and $\op{X}$ within the MSS. Thus, the matrix
representations $\mathsf{A}$ and $\mathsf{B}$ in the basis set
$\{|\tilde{\bm{n}}\rangle_{\rm MSS}\}$ can be constructed by using the algorithm
explained in Alg.~\ref{alg:hmss}. One needs to provide the subroutine that
generates the output states by using the action of the observable operator
$\op{O}$ in a similar way as shown in Alg.~\ref{alg:actH}.
The energy eigenstate expectation values of $\op{O}$ are given by
\begin{equation}
O_\alpha = \langle \alpha | \op{O} | \alpha \rangle =
\sum_{m} S_{m\alpha}^* \left( \sum_n O_{mn} S_{n\alpha} \right) ,
\end{equation}
where $S_{n\alpha}$ is the column vector for the $\alpha$-th eigenstate of
the Hamiltonian~(see Eqs.~\eqref{eigen_eq} and \eqref{simil_S}).

\begin{figure}
\includegraphics*[width=\columnwidth]{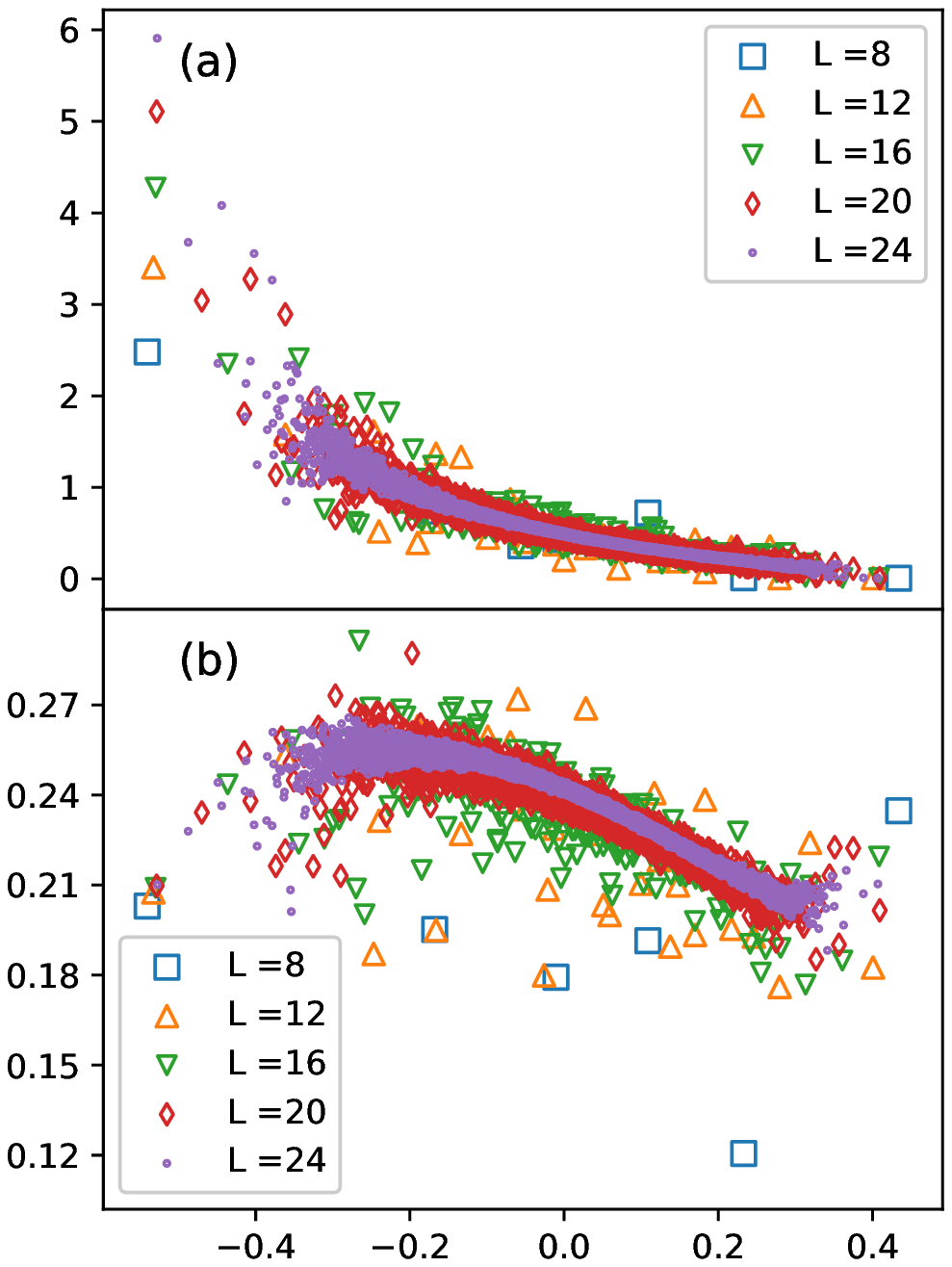}
\caption{Energy eigenstate expectation values of (a) $\op{A}$ and (b) $\op{B}$ 
are plotted as functions of the energy eigenvalue.}\label{fig:Oalpha}
\end{figure}

We plot the eigenstate expectation values in the MSS as 
functions of the energy eigenvalues per site in Fig.~\ref{fig:Oalpha}. 
For a given value of $E_\alpha/L$, the expectation values spread over a finite
range. One can see that the fluctuations become weaker and weaker as $L$ grows. 
The ETH in Eq.~\eqref{ETH_ansatz} predicts that the
fluctuations should scale as $e^{-S(E)}$, which vanishes exponentially as $L$
increases. The numerical data are consistent with the ETH prediction. 
For quantitative analysis of the fluctuations, we refer the readers to
Ref.~\cite{LeBlond:2019bv}.

\subsection{Energy-Temperature Relation}
If the energy eigenstate expectation value depends only on the energy
eigenvalue, the expectation value 
becomes the same as the microcanonical ensemble
average~\cite{DAlessio:2016gr}.
Due to the ensemble equivalence, the
expectation value is then equal to the canonical ensemble average:
\begin{equation}\label{eth_exp}
O_\alpha = {\rm Tr} \op{\rho}_{eq}(\beta) \op{O} ,
\end{equation}
where $\op\rho_{eq}(\beta) = e^{-\beta \op{H}}/Z$ is the density matrix of
the canonical ensemble with the partition function $Z = {\rm Tr} e^{-\beta
\op{H}}$ and the inverse temperature $\beta = 1/ T$.
Consequently, each energy eigenstate $|\alpha\rangle$ with the energy eigenvalue
$E_\alpha$ is assigned to an inverse temperature through the relation
\begin{equation}\label{E_beta}
E_\alpha = {\rm Tr} \op{\rho}_{eq} (\beta) \op{H} 
= \frac{\sum_\gamma E_\gamma e^{-\beta E_\gamma}}{\sum_\gamma
e^{-\beta E_\gamma}} .
\end{equation}

\begin{figure}
\includegraphics*[width=\columnwidth]{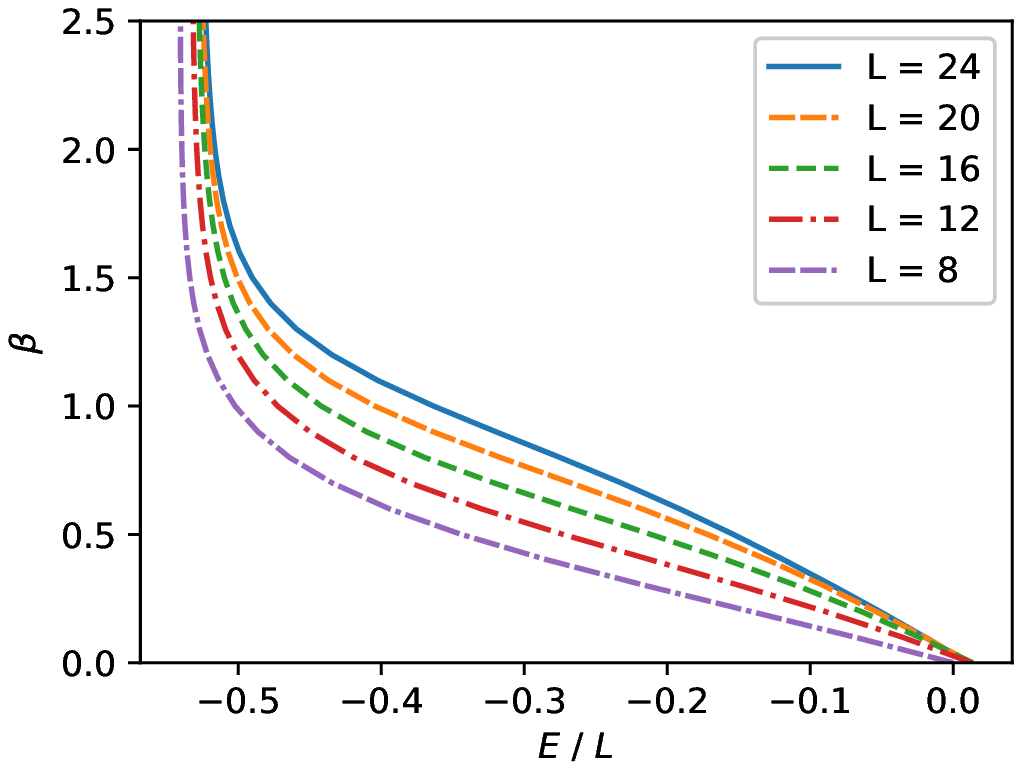}
\caption{Energy density $E/L$ versus inverse temperature $\beta$ 
according to Eq.~\eqref{E_beta}.
}\label{U_beta_curve}
\end{figure}

The energy is an extensive quantity so that the Boltzmann factors $e^{-\beta
E_\alpha}$ are distributed 
widely. In order to reduce possible numerical errors in evaluating the
summation over such quantities, we recommend that 
the Kahan algorithm or the compensation 
algorithm be used~\cite{Kahan:1965:PRR:363707.363723}.
We evaluate numerically the mean energy as a function of $\beta$ in the MSS
for various values of $L$ (Fig.~\ref{U_beta_curve}).

\subsection{Time Evolution}
Suppose that the system is in an initial state 
\begin{equation}
|\psi_0\rangle = \sum_n a_{\bm{n}} |\bm{n}\rangle
\end{equation}
at time $t=0$. Following the Schr\"odinger equation, 
the system evolves into the state 
\begin{equation}\label{schrodinger_sol}
|\psi(t)\rangle = \op{U}(t)|\psi_0\rangle 
\end{equation}
with the time evolution operator
\begin{equation}
\op{U}(t) = \exp[-i \op{H} t] .
\end{equation}
We are interested in the time-dependent 
probability amplitudes $\{a_{\bm{n}}(t)\}$ with which the state 
\begin{equation}
|\psi(t)\rangle = \sum_{\bm{n}} a_{\bm{n}}(t) |\bm{n}\rangle 
\end{equation}
solves the Schr\"odinger equation.

When the whole eigenspectrum of the Hamiltonian operator is available, 
decomposing the initial state in terms of the energy eigenstates 
$\{|\alpha\rangle\}$,
\begin{equation}\label{psi_alpha}
|\psi_0\rangle = \sum_\alpha C_\alpha |\alpha\rangle  ,
\end{equation}
with
\begin{equation}\label{Calpha}
C_\alpha = \langle \alpha | \psi_0\rangle = \sum_{\bm{n}} 
\langle\alpha|\bm{n}\rangle \langle \bm{n} | \psi_0 \rangle = 
\sum_{\bm{n}} S^*_{{\bm{n}}\alpha} a_{\bm{n}}  ,
\end{equation}
is convenient. Note that $\mathsf{S}=\{S_{{\bm{n}}\alpha}\}$ is the unitary matrix 
diagonalizing $\mathsf{H}$~(see Eq.~\eqref{simil_S}). 
Inserting Eq.~\eqref{psi_alpha} into Eq.~\eqref{schrodinger_sol}, 
one obtains
\begin{equation}\label{ant}
a_{\bm{n}}(t) = \sum_\alpha S_{{\bm{n}}\alpha} 
\left(C_\alpha e^{-iE_\alpha t}\right)  .
\end{equation}
Thus, the state at arbitrary time $t$ can be found 
by performing the matrix-vector multiplication in 
Eqs.~\eqref{Calpha} and
\eqref{ant}. If the initial state $|\psi_0\rangle$ belongs to a specific
symmetry sector, knowledge of the eigenvalues and the eigenvectors
within the sector is sufficient.

\begin{figure}
\includegraphics*[width=\columnwidth]{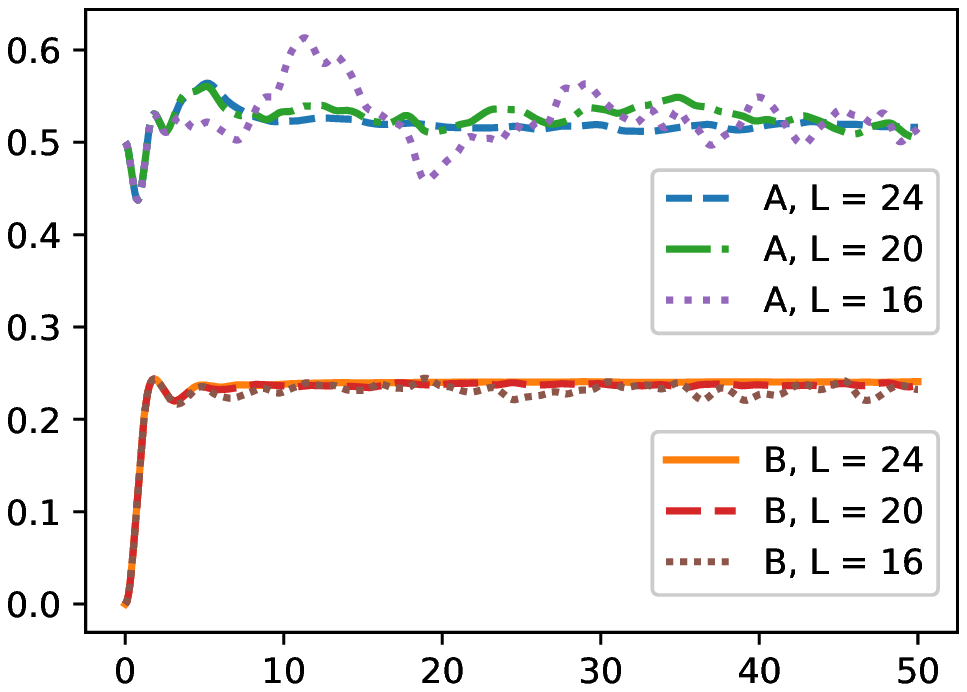}
\caption{Time-dependent expectation values of $\op{A}$ and
$\op{B}$ for systems with $L=16, 20$, and $24$.}\label{fig:t_dep}
\end{figure}

We demonstrate the time evolution starting from the initial state
\begin{equation}\label{psi0}
|\psi_0\rangle = \frac{1}{\sqrt{2}} ( |0101\cdots\rangle +
|1010\cdots\rangle) .
\end{equation} 
It belongs to the MSS with SRS
$|\tilde{\bm{n}}\rangle = |0101\cdots\rangle$. It is decomposed into
a linear superposition of the energy eigenstates in the MSS; then, the
probability amplitudes at time $t$ are obtained from Eq.~\eqref{ant}. In
Fig.~\ref{fig:t_dep}, we present the time-dependent expectation values of
$\op{A}$ and $\op{B}$. After an initial transient region, both quantities
relax into stationary values with fluctuations. The amplitude of
the fluctuations decreases as $L$ increases for both quantities.
An approach to a stationary state is guaranteed when the initial state overlaps
a large enough number of energy
eigenstates~\cite{Reimann:2008hq,Short:2011fq}.

When the whole set of eigenvalues and the eigenvectors is not 
available, the Lie-Trotter-Suzuki~(LTS) decomposition method is
useful~\cite{Suzuki:1985fz}. 
The Hamiltonian is a sum of local operators that do not commute 
with one another. 
Rearranging those local operators, one can decompose
the Hamiltonian into multiple partial Hamiltonians in such a way that
each partial Hamiltonian should consist of a set of mutually commuting local 
operators. 
For clarity, we explain the method with the XXZ Hamiltonian in
Eq.~\eqref{HXXZ} with only the nearest neighbor interaction~(see
Eq.~\eqref{ults_nnn} for the case with the next-nearest neighbor
interactions).
The Hamiltonian can be written as 
\begin{equation}
\op{H} = \sum_{l=0}^{L-1} \op{h}_{l,l+1} , 
\end{equation}
where 
\begin{equation}
\op{h}_{l,m} = -\op{\sigma}_l^+ \op{\sigma}_m^- - \op{\sigma}_l^-\op\sigma_m^+
- \frac{\Delta}{2}\op\sigma_l^z \op\sigma_m^z
\end{equation}
is the XXZ coupling between two spins at sites $l$
and $m$.
The Hamiltonian can be written as 
\begin{equation}
\op{H} = \op{H}_0+\op{H}_1
\end{equation}
with $\op{H}_0 = \sum_{l=0}^{L/2-1} \op{h}_{2l,2l+1}$
and $\op{H}_1 = \sum_{l=0}^{L/2-1} \op{h}_{2l+1,2l+2}$. 
Checking that $\op{H}_0$ and $\op{H}_1$ are the sum of 
mutually commuting operators, respectively, is easy.

The LTS method is based on the Baker-Campbell-Hausdorff formula
\begin{equation}
e^{\delta(\op{A}+\op{B})} = e^{\delta \op{A}} e^{\delta \op{B}}
e^{-\frac{\delta^2}{2} [\op{A},\op{B}]} \cdots 
\end{equation}
Applying it to $\op{U}(t=\epsilon)=e^{-i\epsilon (\op{H}_0 + 
\op{H}_1)}$, 
one obtains  $\op{U}(\epsilon) =
\op{U}_{LTS}(\epsilon) + O(\epsilon^2)$ with the approximate
time evolution operator 
\begin{equation}
\op{U}_{LTS}(\epsilon) = e^{-i \epsilon \op{H}_0} e^{-i \epsilon \op{H}_1}  .
\end{equation}
Due to the commutation property, $\op{U}_{LTS}(\epsilon)$ takes the
product form
\begin{equation}
\op{U}_{LTS}(\epsilon) = \left[ \prod_{l=0}^{L/2-1} \op{u}_{2l,2l+1}\right]
\left[ \prod_{l=0}^{L/2-1} \op{u}_{2l+1,2l+2} \right] ,
\end{equation}
where
\begin{equation}
\op{u}_{l,m} \equiv e^{ -i\epsilon \op{h}_{l,m}}
\end{equation}
acts only on two sites $l$ and $m$. 
The true time evolution operator $\op{U}(\epsilon)$ rotates all spins
simultaneously. On the other hand, $\op{U}_{LTS}(\epsilon)$ 
rotates pairs of spins successively, which is easily done computationally.
The two-spin rotation operator $\op{u}_{l,m}$ is represented by the 
$4\times 4$ matrix
\begin{equation}
\mathsf{u} = \begin{pmatrix}
e^{i\Delta \epsilon /2} & 0 & 0 & 0 \\
0 & e^{-i \Delta \epsilon/2} \cos\epsilon & ie^{-i \Delta \epsilon/2} \sin{\epsilon} & 0 \\
0 & ie^{-i \Delta \epsilon/2} \sin\epsilon & e^{-i \Delta \epsilon/2} \cos{\epsilon} & 0 \\
0 & 0 & 0 & e^{i \Delta \epsilon/2}
\end{pmatrix}
\end{equation}
with the basis set 
$$ 
\{ |0\rangle_l\otimes|0\rangle_m, |0\rangle_l\otimes|1\rangle_m,
|1\rangle_l\otimes|0\rangle_m,|1\rangle_l\otimes|1\rangle_m\} .
$$
The pseudocode for the two-spin rotation is presented in
Alg.~\ref{alg:rotation2}.
\begin{figure}
\begin{algorithm}[H]
\caption{Two-spin rotation $|\psi'\rangle = \op{u}_{l,m} |\psi\rangle$ 
for $|\psi\rangle = \sum_{\bm{n}} a_{\bm{n}}
|{\bm{n}}\rangle$}\label{alg:rotation2}
\begin{algorithmic}
\Procedure{Acting\_u}{$|\psi\rangle,L,l,m$}
\State{$a_{\bm{n}}' = 0$ for all $\bm{n}$}
\For{(${\bm{n}}=0$ to $2^L-1$)}
   \State{$n_l = l\mbox{th bit of } n$ ; $n_m = m\mbox{th bit of } {\bm{n}}$}
   \If {$(n_l = n_m)$}
      \State{$a'_{\bm{n}} = a'_{\bm{n}} + a_{\bm{n}} e^{i \Delta \epsilon /2}$}
   \Else
      \State{$a'_{\bm{n}} = a'_{\bm{n}} + a_{\bm{n}} e^{-i \Delta \epsilon} \cos\epsilon$}
      \State{${\bm{m}} = \mbox{bit\_flip}(\bm{n},l,m)$}
      \State{$a'_{{\bm{m}}} = a'_{{\bm{m}}} + a_{\bm{n}} i e^{-i\Delta \epsilon/2}
                \sin\epsilon$}
   \EndIf
\EndFor
\State{Return( $|\psi'\rangle = \sum_{\bm{n}} a'_{\bm{n}} |{\bm{n}}\rangle$ )}
\EndProcedure
\end{algorithmic}
\end{algorithm}
\end{figure}

A few remarks are in order: (i) One may consider the approximation
$\op{U}(\epsilon) \simeq 1 - i\epsilon \op{H} +O(\epsilon^2)$, 
which is analogous to the
Euler method for ordinary differential equations. Such an approximation
is not recommended because it breaks unitarity of the time evolution
operator~\cite{Askar:1978fq}. 
(ii) $\op{U}_{LTS}(\epsilon)$ is unitary, 
conserves the particle number, and commutes with
$\op{X}$ as the original time evolution operator $\op{U}$. 
Unfortunately, however, it does not commute with $\op{T}$ and $\op{R}$.
If the symmetry property is important, one should  use
the symmetrized form 
$\op{U}_{LTS,s}(\epsilon) = (e^{-i \epsilon \op{H}_0} e^{-i\epsilon \op{H}_1}
 +e^{-i \epsilon \op{H}_1} e^{-i\epsilon \op{H}_0})/2$.
(iii) The wave function after time $t$ is obtained by applying the
infinitesimal time evolution operator $t/\epsilon$ times, which leads to a numerical
error of $O(t \epsilon)$. If one uses the higher-order expansion
formula~\cite{Suzuki:1985fz}
\begin{equation}
e^{\delta (\op{A}+\op{B})} = e^{\frac{\delta}{2} \op{A}} e^{\delta \op{B}}
e^{\frac{\delta}{2} \op{A}} + O(\delta^3) \ ,
\end{equation}
one can reduce the numerical error. Thus, using
the higher order approximation
\begin{equation}\label{higherorder}
\op{U}_{LTS}(\epsilon) =  e^{-i\epsilon \op{H}_0/2} e^{-i\epsilon\op{H}_1}
e^{-i\epsilon\op{H}_0/2} \ ,
\end{equation}
whose overall numerical error scales as $O(t\epsilon^2)$, would be wise.

When the Hamiltonian
includes next-nearest neighbor interactions as in Eq.~\eqref{Hnnn}, 
the Hamiltonian may be decomposed as 
\begin{equation}
\op{H} = \op{H}_0 + \left( \op{H}_1 + \left(\op{H}_2 +
\op{H}_3\right)\right)
\end{equation}
with $\op{H}_2 =
\sum_{l=0}^{L/4-1}(\op{h}_{4l,4l+2}+\op{h}_{4l+1,4l+3})$ and $\op{H}_3 =
\sum_{l=0}^{L/4-1}(\op{h}_{4l+2,4l+4}+\op{h}_{4l+3,4l+5})$. 
Applying Eq.~\eqref{higherorder} successively~\cite{Noh:2019gx}, one
find that $\op{U}(\epsilon) = \op{U}_{LTS}(\epsilon) + O(\epsilon^3)$, where
\begin{equation}\label{ults_nnn}
\begin{split}
\op{U}_{LTS}(\epsilon) =&
e^{-i\epsilon\op{H}_0/2}e^{-i\epsilon\op{H}_1/2}e^{-i\epsilon\op{H}_2/2} 
 e^{-i\epsilon\op{H}_3} \\
& \times
e^{-i\epsilon\op{H}_2/2}e^{-i\epsilon\op{H}_1/2}e^{-i\epsilon\op{H}_0/2} .
\end{split}
\end{equation}

\begin{figure}
\includegraphics*[width=\columnwidth]{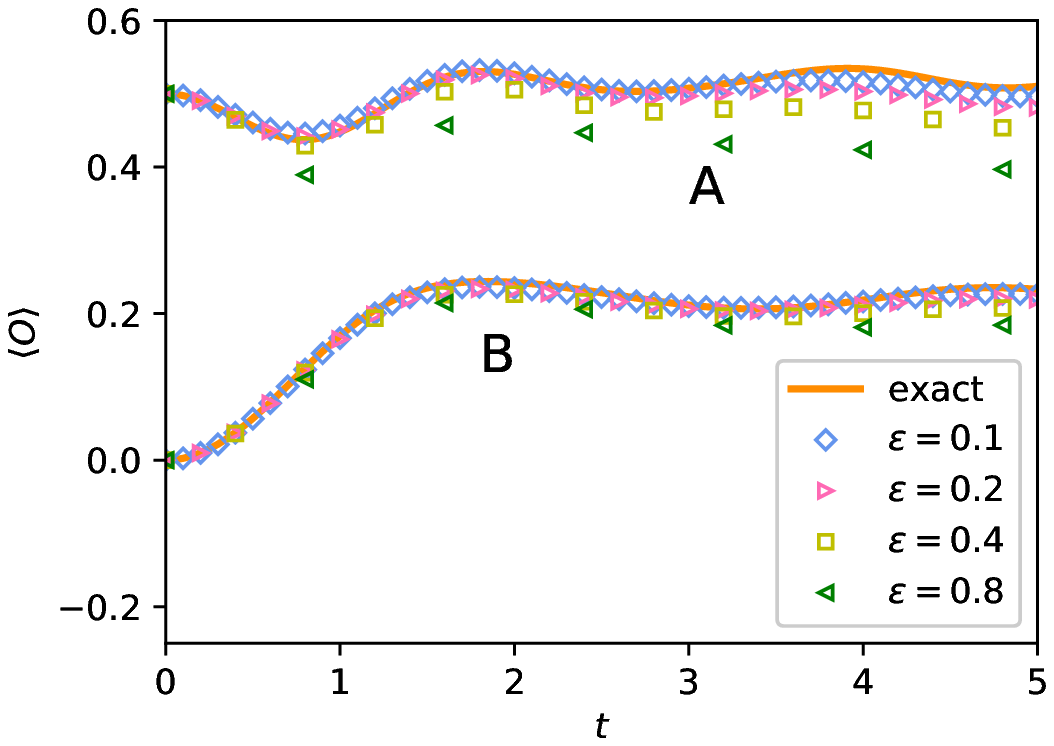}
\caption{Time evolution of the expectation values of $\op{O} = \op{A}$ and 
$\op{B}$ calculated by using the exact time-dependent wave function 
and the approximate time-evolution operator in Eq.~\eqref{ults_nnn} with
$\epsilon=0.1,\cdots,0.8$. The lattice size is $L=14$.}
\label{fig:LTS}
\end{figure}

We compare in Fig.~\ref{fig:LTS} the time evolutions of the expectation values of $\op{A}$ and
$\op{B}$ calculated from the exact wave function in
Eq.~\eqref{ant} and from the approximate decomposition method in
Eq.~\eqref{ults_nnn}. The lattice size is $L=14$, and the
initial state is $|\psi_0\rangle$ in Eq.~\eqref{psi0}.
As $t$ increases, the approximate solution deviates from the exact solution.
The error decreases as $\epsilon$ becomes smaller. Because the infinitesimal
time evolution operator has an error of $O(\epsilon^3)$, the numerical error at
finite $t$ scales as $O(t\epsilon^2)$.

\subsection{Entanglement Entropy}
The entanglement is a characteristic feature of a quantum mechanical
system~\cite{Amico:2008en,Kim:2013fy,Vidmar:2018el}.
It measures the extent to which a part $S_1$ of a system and its complement
$S_2 = S_1^c$ are interwound with each other. Let $\{|\bm{r}\rangle\}$ and
$\{|\bm{l}\rangle\}$ be the basis sets of the subsystems $S_1$ and $S_2$ 
whose Hilbert space dimensions are $D_1$ and $D_2$, respectively. 
Then, a state vector can be written as
\begin{equation}\label{direct_product}
|\psi\rangle = \sum_{\bm{l},\bm{r}} \psi_{\bm{l}\bm{r}}
|\bm{l}\rangle\otimes|\bm{r}\rangle .
\end{equation}
If the probability amplitudes are factorized as
$\psi_{\bm{l}\bm{r}} = \phi_{\bm{l}} \varphi_{\bm{r}}$, the state vector is
unentangled and is given by the direct product $|\psi\rangle = 
\left(\sum_{\bm{l}} \phi_{\bm{l}}|\bm{l}\rangle\right)\otimes 
\left(\sum_{\bm{r}} \varphi_{\bm{r}}|\bm{r}\rangle\right)$. 
Otherwise, it is entangled. 
The entanglement can be quantified by using the von Neumann entropy
\begin{equation}
S_E = - {\rm Tr}_1 \op\rho_1 \ln \op\rho_1
\end{equation}
of the reduced density matrix 
\begin{equation}
\op\rho_1 = {\rm Tr}_{2} |\psi\rangle\langle\psi| = 
\sum_{\bm{l}} \psi_{\bm{l}\bm{r}}
\psi_{\bm{l}\bm{r}'}^* |\bm{r}\rangle \langle \bm{r}'| 
\end{equation}
for the subsystem $S_1$.
In terms of the eigenvalues $\{\lambda^1_i\}$ of $\op\rho_1$,  
the entanglement entropy is given by
\begin{equation}
S_E = -\sum_{i=1}^{D_1} \lambda^1_i \ln \lambda^1_i \ .
\end{equation}

The singular value decomposition~(SVD) is extremely useful in calculating the
reduced density matrix and the entanglement entropy.
The probability amplitude $\psi_{\bm{lr}}$ can be regarded as an element of 
the $D_2\times D_1$ matrix $\mathsf{\Psi}$. 
Any matrix with complex elements can be written in a product
form~\cite{Golub:1996wp}
\begin{equation}
\mathsf{\Psi} = \mathsf{U} \mathsf{\Sigma} \mathsf{V}^\dagger  ,
\end{equation}
with a $D_2\times D_1$ rectangular matrix $\mathsf{\Sigma}$ and 
a unitary matrix $\mathsf{U}~(\mathsf{V})$ of size
$D_{\rm 2}\times D_{\rm 2}~(D_{\rm 1}\times D_{\rm 1})$.
The off-diagonal elements  of the rectangular matrix $\Sigma$ are zero, and 
the diagonal elements $\{\Sigma_1,\cdots, \Sigma_{\min[D_{\rm 2},D_{\rm 1}]}\}$
are real and nonnegative~\cite{Golub:1996wp}. 
The diagonal elements are called the singular values. 
The reduced density matrix for the subsystem $S_1$ is given by
\begin{equation}
\mathsf{\rho}_1 = \mathsf{\Psi}^\dagger\mathsf{\Psi} = \mathsf{V} 
(\mathsf{\Sigma}^\dagger\mathsf{\Sigma}) \mathsf{V}^\dagger ,
\end{equation}
which is the similarity transformation of the $(D_1\times D_1)$ diagonal matrix 
$\mathsf{\Sigma}^\dagger\mathsf{\Sigma} = {\rm diag}[\Sigma_1^2,\cdots,
\Sigma_{\min[D_1,D_2]}^2,0,\cdots,0]$.
Thus, the entanglement entropy is given by
\begin{equation}
S_E = -\sum_{i=1}^{\min[D_1,D_2]} (\Sigma_i)^2 \ln (\Sigma_i)^2 .
\end{equation}
The reduced density matrix for the alternative subsystem $S_2$ is given by
$\rho_2 = \mathsf{\Psi}\mathsf{\Psi}^\dagger =
\mathsf{U(\Sigma\Sigma}^\dagger)\mathsf{U}^\dagger$. It shares the same
nonzero eigenvalues with $\rho_1$. 
Thus, the two subsystems $S_1$ and $S_2$ have
the same von Neumann entropy and yield the same entanglement entropy.

Using the SVD method, one can calculate the entanglement entropy of the
one-dimensional system efficiently.
Consider a partition of the one-dimensional lattice of $L$ sites into two
subsets: $S_1$ for the rightmost $L_{\rm 1}$ sites at 
$l=0,1,\cdots,L_1-1$ and $S_2$ for
$L_2 = (L-L_1)$ sites at $l=L_1,\cdots,L-1$.
A basis state  $|\bm{n}\rangle$ with $0\leq \bm{n} < 2^L$ 
for the whole system can be written as a product
state $|\bm{l}\rangle\otimes|\bm{r}\rangle$, where $0 \leq \bm{l} <
D_2=2^{L_2}$ and $0\leq \bm{r}< D_1=2^{L_1}$ are
related to $\bm{n}$ through
\begin{equation}\label{rem_quo}
\bm{n} = 2^{L_1} \bm{l} + \bm{r}  .
\end{equation}
That is, $\bm{r}$ and $\bm{l}$ are the remainder and the quotient in the
integer division of $\bm{n}$ by $D_1=2^{L_1}$.
Accordingly, any
state $|\psi\rangle=\sum_{\bm{n}} a_{\bm{n}} | \bm{n}\rangle$ can be written 
in the form of Eq.~\eqref{direct_product} by identifying
$\psi_{\bm{lr}}=a_{\bm{n}}$. If one is interested only in the
entanglement entropy, calculating the singular values while neglecting
the unitary matrices $\mathsf{U}$ and $\mathsf{V}$ will suffice.
A pseudocode for the entanglement entropy is presented in Alg.~\ref{alg7}.
The crucial part is the subroutine to perform the SVD, which is found in 
standard numerical libraries.

\begin{figure}
\begin{algorithm}[H]
\caption{Entanglement entropy for a state $|\psi\rangle = \sum_{\bm{n}} 
a_{\bm{n}} |{\bm{n}}\rangle$}\label{alg7}
\begin{algorithmic}
\Procedure{EntanglementEntropy}{$|\psi\rangle,L,L_1,L_2$}
\For{(${\bm{n}}=0$ to $2^L-1$)}
   \State{$\bm{r} = \mbox{remainder of }\bm{n} / 2^{L_1}$;  
    $\bm{l} = \mbox{quotient of }\bm{n} / 2^{L_1}$ }
   \State{$ \psi_{\bm{l}\bm{r}} = a_{\bm{n}} $}
\EndFor

\State{$(\mathsf{\Sigma,U,V}) \gets
\mbox{\textsc{SVD}}(\mathsf{\Psi},2^{L_2},2^{L_1})$}
\State{${\tt Entropy} = - \sum_i (\Sigma_i)^2 \ln (\Sigma_i)^2$}
\State{Return({\tt Entropy})}
\EndProcedure
\end{algorithmic}
\end{algorithm}
\end{figure}

\begin{figure}
\includegraphics*[width=\columnwidth]{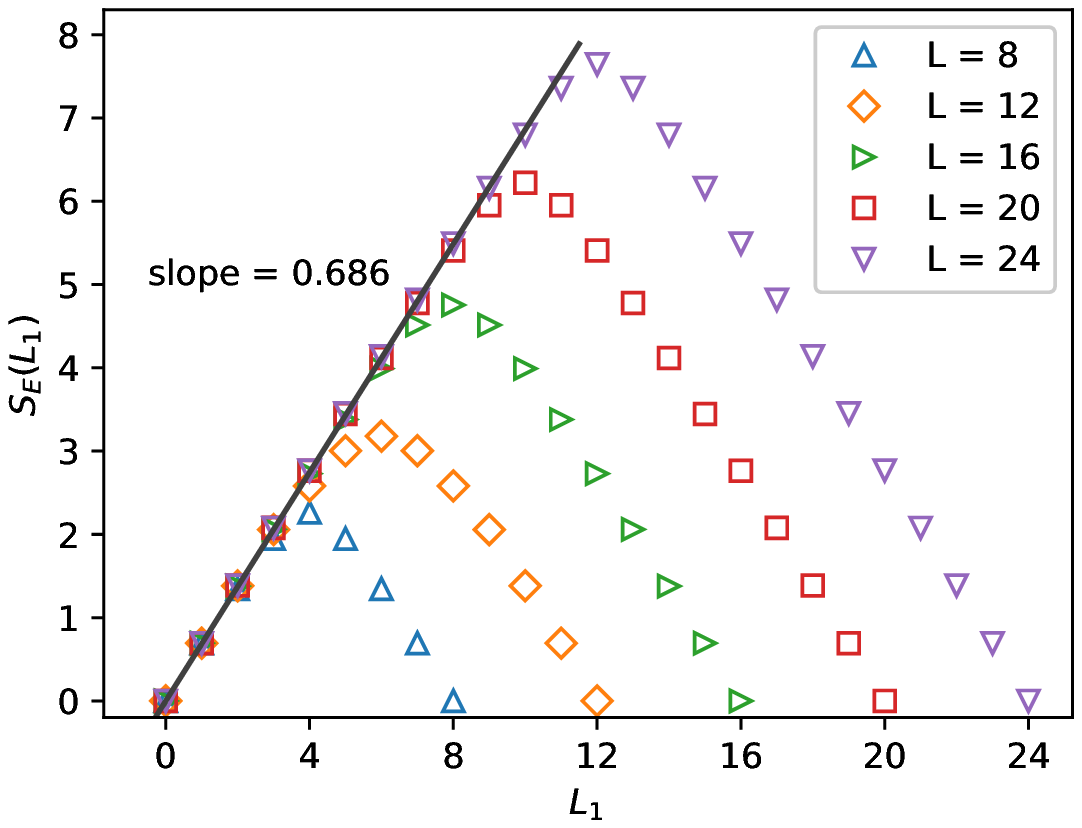}
\includegraphics*[width=\columnwidth]{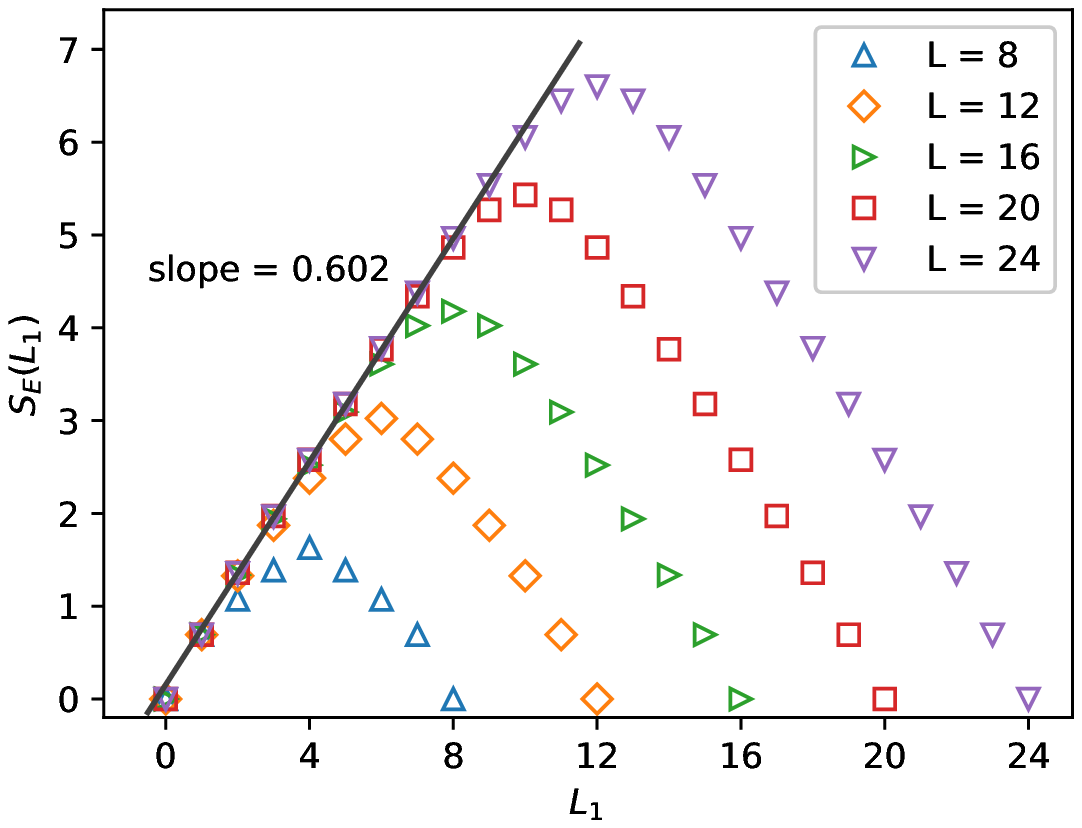}
\caption{Entanglement entropy $S_E(L_R)$ of the energy eigenstates whose 
energy densities are closest to $0$ and $-0.2$.}
\label{fig:S_E}
\end{figure}

We have calculated the entanglement entropy for
the energy eigenstate of the XXZ Hamiltonian in Eq.~\eqref{Hnnn}. For each
$L$, we have chosen the eigenstate $|\alpha\rangle$ in the MSS whose energy
density $E_\alpha/L$ is closest to $-0.2$ or $0$. The entanglement entropy
$S_E(L_1)$ as a function of the subsystem size $L_1$ is shown 
in Fig.~\ref{fig:S_E}. The von Neumann entropy of the subsystems $S_1$ and
$S_2$ are the same. 
Thus, $S_E(L_1) = S_E(L-L_1)$ and $S_E(0) =S_E(L)=0$. For $0\ll
L_1 \ll L/2$, the entanglement entropy is proportional to $L_1$, obeying 
the volume law~\cite{Garrison:2018hi}. 
Numerical estimates for the slope are $s =
0.60$ and $s=0.69$ for the energy densities $E_\alpha/L = -0.2$ and $0.0$,
respectively.
For quantum systems that thermalize, the slope is equal to the 
thermodynamic entropy density $s = S/L$~\cite{Garrison:2018hi}. 
The system with $E_\alpha/L=0.0$ is in an infinite temperature 
state~(see Fig.~\ref{U_beta_curve}) where the thermodynamic entropy 
density is $s=\ln 2$. The numerical result for the slope $s=0.69$ for the
states with $E_\alpha/L=0.0$ is consistent with the entropy density at 
infinite temperature.

\begin{figure}
\includegraphics*[width=\columnwidth]{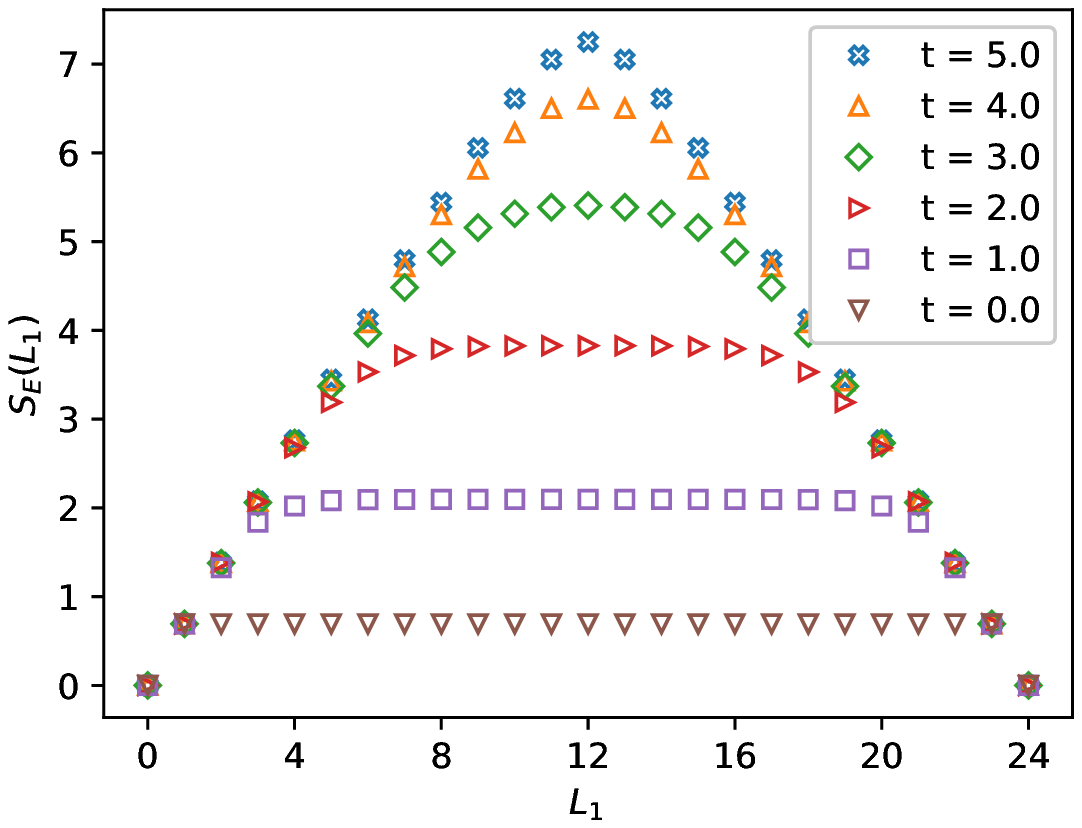}
\caption{Time evolution of the entanglement entropy for the system with
$L=24$.}
\label{fig:SEt}
\end{figure}

We also study the time evolution of the entanglement entropy starting from the
initial state $|\psi_0\rangle$ in Eq.~\eqref{psi0}, which is not the energy
eigenstate.
Figure~\ref{fig:SEt} shows the numerical results obtained for system 
with $L=24$. The initial state has the reduced density matrix 
$$
\op{\rho}_1 = \frac{1}{2} \left( |\overbrace{01\cdots}^{L_1}\rangle\langle
\overbrace{01\cdots}^{L_1}| +
|\overbrace{10\cdots}^{L_1}\rangle\langle \overbrace{10\cdots}^{L_1}|\right) ,
$$
whose von Neumann entropy is $\ln 2$ for $L_1 \neq 0, L$. 
The plateau at $t=0$ explains the initial entanglement entropy.
Linear regions appear near $L_1 = 0$ and $L$ at $t>0$, the size of
which grows in time. Eventually, the entanglement entropy converges to a
stationary distribution that satisfies the volume law~\cite{Kim:2013fy}.

\section{Summary}\label{sec:summary}
In this paper, we present a thorough review of a numerical method to
diagonalize the quantum mechanical Hamiltonian. The method is applicable to
systems with finite Hilbert space dimensionality. In the presence of 
particle number conservation and translational invariance, the Hamiltonian
matrix can be broken up into symmetry sectors specified with the eigenvalues
of the number operator and the shift operator. One can further make use of 
discrete symmetry such as the particle-hole symmetry and the spatial inversion
symmetry to reduce the matrix size effectively.

As an application of the numerical method, we studied the XXZ model 
Hamiltonian with nearest and next-nearest neighbor interactions. We present
various numerical results in the maximum symmetry sector up to the lattice
size $L=24$. The energy eigenvalues~(Fig.~\ref{E_L8}) and the CPU time for the
diagonalization~(Fig.~\ref{fig:cpu_time}) are presented in Section~\ref{sec:nd}.
The XXZ Hamiltonian is a prototypical model for the study of quantum
thermalization. We also present the expectation values of
observables~(Fig.~\ref{fig:Oalpha}), the inverse
temperature~(Fig.~\ref{U_beta_curve}), the relaxation dynamics, and the 
entanglement entropy~(Fig.~\ref{fig:S_E}) in Section~\ref{sec:thermal} with the
numerical methods for those quantities. We hope that our review will be helpful to
those who are interested in the numerical study of the quantum thermalization 
of isolated quantum systems.

\section*{ACKNOWLEDGEMENT}
This work is supported by the 2016 Research Fund of the University of Seoul.


\bibliography{paper}

\appendix
\begin{center} {\bf Appendix} \end{center}
\begin{widetext}
\setcounter{MaxMatrixCols}{20}
In this Appendix, we present the explicit representation of the Hamiltonian
matrix for systems with small $L$.
When $L=4$, the Hamiltonian in Eq.~\eqref{HXXZ} can be represented by a 
$16\times 16$ matrix in the basis set 
$\{|\bm{0}\rangle,\cdots,|\bm{15}\rangle\}$, given by
\begin{equation}\label{H4full}
\mathsf{H} = \begin{pmatrix}
 -2\Delta &  0 &  0 &  0 &  0 &  0 &  0 &  0 &  0 &  0 &  0 &  0 &  0 &  0 &  0 &  0\\
 0 &  0 & -1 &  0 &  0 &  0 &  0 &  0 & -1 &  0 &  0 &  0 &  0 &  0 &  0 &  0\\
 0 & -1 &  0 &  0 & -1 &  0 &  0 &  0 &  0 &  0 &  0 &  0 &  0 &  0 &  0 &  0\\
 0 &  0 &  0 &  0 &  0 & -1 &  0 &  0 &  0 &  0 & -1 &  0 &  0 &  0 &  0 &  0\\
 0 &  0 & -1 &  0 &  0 &  0 &  0 &  0 & -1 &  0 &  0 &  0 &  0 &  0 &  0 &  0\\
 0 &  0 &  0 & -1 &  0 &  2\Delta & -1 &  0 &  0 & -1 &  0 &  0 & -1 &  0 &  0 &  0\\
 0 &  0 &  0 &  0 &  0 & -1 &  0 &  0 &  0 &  0 & -1 &  0 &  0 &  0 &  0 &  0\\
 0 &  0 &  0 &  0 &  0 &  0 &  0 &  0 &  0 &  0 &  0 & -1 &  0 &  0 & -1 &  0\\
 0 & -1 &  0 &  0 & -1 &  0 &  0 &  0 &  0 &  0 &  0 &  0 &  0 &  0 &  0 &  0\\
 0 &  0 &  0 &  0 &  0 & -1 &  0 &  0 &  0 &  0 & -1 &  0 &  0 &  0 &  0 &  0\\
 0 &  0 &  0 & -1 &  0 &  0 & -1 &  0 &  0 & -1 &  2\Delta &  0 & -1 &  0 &  0 &  0\\
 0 &  0 &  0 &  0 &  0 &  0 &  0 & -1 &  0 &  0 &  0 &  0 &  0 & -1 &  0 &  0\\
 0 &  0 &  0 &  0 &  0 & -1 &  0 &  0 &  0 &  0 & -1 &  0 &  0 &  0 &  0 &  0\\
 0 &  0 &  0 &  0 &  0 &  0 &  0 &  0 &  0 &  0 &  0 & -1 &  0 &  0 & -1 &  0\\
 0 &  0 &  0 &  0 &  0 &  0 &  0 & -1 &  0 &  0 &  0 &  0 &  0 & -1 &  0 &  0\\
 0 &  0 &  0 &  0 &  0 &  0 &  0 &  0 &  0 &  0 &  0 &  0 &  0 &  0 &  0 & -2\Delta
\end{pmatrix}
\end{equation}
Using particle number conservation, we can write the Hamiltonian matrix in
Eq.~\eqref{H4full} 
in block-diagonal form as
\begin{equation}\label{H4N}
\mathsf{H} = \begin{pmatrix}
 \boxed{\begin{matrix} -2\Delta \end{matrix}} & & & & \\
  & \boxed{\begin{matrix}
            0 & -1 & 0 & -1 \\
            -1 & 0 & -1 & 0 \\
            0 & -1 & 0 & -1 \\
            -1 & 0 & -1 & 0 \\
     \end{matrix}} & & &  \\
 & & \boxed{\begin{matrix}
             0 & -1 & 0 & 0 & -1 & 0 \\
             -1 & 2\Delta & -1 & -1 & 0 & -1 \\
             0 & -1 & 0 & 0 & -1 & 0 \\
             0 & -1 & 0 & 0 & -1 & 0 \\
             -1 & 0 & -1 & -1 & 2\Delta & -1  \\
             0 & -1 & 0 & 0 & -1 & 0 
     \end{matrix}} & & \\
 & & & \boxed{\begin{matrix}
             0 & -1 & 0 & -1  \\
             -1 & 0 & -1 & 0  \\
             0 & -1 & 0 & -1  \\
             -1 & 0 & -1 & 0 
     \end{matrix}} &  \\
 & & & & \boxed{\begin{matrix} -2\Delta \end{matrix}}
\end{pmatrix} .
\end{equation}
\end{widetext}
Each block corresponds to an $N$-particle sector with $N=0~(\mbox{top}), 1,
2, 3, 4~(\mbox{bottom})$. 
Using the translational symmetry, we can further block-diagonalize
$\mathsf{H}_N$ by using the wave-number quantum number $k=0,\cdots,L-1$. The
block-diagonal form of $\mathsf{H}_2$ in the $N=2$ particle sector is given
by
\begin{equation}\label{H4N2k}
\mathsf{H}_2 = \begin{pmatrix}
 \boxed{\begin{matrix} 0 & -2\sqrt{2} \\ -2\sqrt{2} & 2\Delta \end{matrix}} & & & \\
  & \boxed{\begin{matrix} 0 \end{matrix}} & & \\
  & & \boxed{\begin{matrix} 0 & 0 \\ 0 & 2\Delta \end{matrix}} & \\
  & & & \boxed{\begin{matrix} 0 \end{matrix}} 
\end{pmatrix} \end{equation}  
where the blocks correspond to $\mathsf{H}_{2,0}$~(top), $\mathsf{H}_{2,1}$,
$\mathsf{H}_{2,2}$, and $\mathsf{H}_{2,3}$~(bottom).
The Hamiltonian matrix in the MSS for $L=8$ is given by
\begin{equation}\label{HmssL8}
\mathsf{H}_{\rm MMS} = \begin{pmatrix}
-2\Delta  & -\sqrt{2} &  0 & 0          &   0 &  0         &0 \\
-\sqrt{2} & 0         & -2 & -\sqrt{2}  &   0 &  0         &0  \\
0         & -2        &  0 & 0          &   0 & -2         &0 \\
0         & -\sqrt{2} &  0 & 2\Delta    &   0 & -2\sqrt{2} &0 \\
0         & 0         &  0 & 0          &   0 & -2         &0 \\
0         & 0         & -2 & -2\sqrt{2} &  -2 &  2\Delta   &-2\sqrt{2} \\
0         & 0         &  0 & 0          &   0 & -2\sqrt{2} &4\Delta 
\end{pmatrix}
\end{equation}

\end{document}